\documentclass[preprint,12pt]{elsarticle}

\usepackage{amssymb}
\usepackage{amsmath}
\usepackage{booktabs}
\usepackage{threeparttable}
\usepackage{float}
\usepackage{makecell}
\usepackage{needspace}
\usepackage{array}
\usepackage{xurl}
\usepackage{xcolor}
\usepackage[hidelinks]{hyperref}

\definecolor{revisionyellow}{RGB}{0,0,0}
\newcommand{\rev}[1]{{#1}}
\newenvironment{revisionblock}{\begingroup}{\endgroup}

\journal{Energy Storage Materials}

\begin{document}

\hbadness=10000
\vbadness=10000
\hfuzz=4pt
\emergencystretch=3em

\begin{frontmatter}

\title{When voltage sensors fail: Electrochemically constrained fault-tolerant state estimation for flat-plateau LFP batteries\tnoteref{pubnote}}

\tnotetext[pubnote]{This is the authors' accepted manuscript of the article published in \textit{Energy Storage Materials}, Volume 91 (2026), Article 105553. Please cite the Version of Record: \url{https://doi.org/10.1016/j.ensm.2026.105553}. Official published version: \url{https://www.sciencedirect.com/science/article/pii/S2405829726006859}. \textcopyright\ 2026. This manuscript version is made available under the CC BY-NC-ND 4.0 license: \url{https://creativecommons.org/licenses/by-nc-nd/4.0/}.}

\author[inst1,inst2,inst3]{Feng Guo}
\author[inst2,inst3]{Luis D. Couto}
\author[inst1,inst3]{Hamid Hamed}
\author[inst2,inst3]{Khiem Trad}
\author[inst4]{Dong Zhang}
\author[inst5]{Ru Hong}
\author[inst6]{Guangdi Hu}
\author[inst1,inst3]{Mohammadhosein Safari\corref{cor1}}
\ead{momo.safari@uhasselt.be}

\cortext[cor1]{Corresponding author.}

\affiliation[inst1]{organization={Institute for Materials Research (IUMAT), Hasselt University},
    addressline={Martelarenlaan 42},
    city={Hasselt},
    postcode={3500},
    country={Belgium}}

\affiliation[inst2]{organization={Water and Energy Transition, VITO},
    addressline={Boeretang 200},
    city={Mol},
    postcode={2400},
    country={Belgium}}

\affiliation[inst3]{organization={EnergyVille},
    addressline={Thor Park 8310},
    city={Genk},
    postcode={3600},
    country={Belgium}}

\affiliation[inst4]{organization={School of Aerospace and Mechanical Engineering, The University of Oklahoma},
    city={Norman},
    postcode={OK 73019},
    country={USA}}

\affiliation[inst5]{organization={College of Tianyin Lake Science and Technology Innovation, Nanjing Institute of Technology},
    city={Nanjing},
    postcode={211167},
    country={China}}

\affiliation[inst6]{organization={School of Vehicles and Intelligent Transportation, Fuyao University of Science and Technology},
    city={Fuzhou},
    postcode={350109},
    country={China}}

\begin{abstract}
\begin{revisionblock}
The flat voltage plateau of lithium iron phosphate (LFP)/graphite cells makes electrochemical-state errors and voltage-measurement abnormalities produce similar innovations, complicating state-of-charge (SOC) estimation. This work proposes an electrochemically constrained residual-bias compensation dual extended Kalman filter (RBC-DEKF) with uncertain-initialization commissioning. A thermal control-oriented parameter-grouped single-particle model (CPG-SPMT) provides paired-electrode dynamics and terminal-voltage prediction, while a separately configurable voltage-discrepancy state accommodates systematic residuals. When the initial SOC is uncertain, residual adaptation is suspended over a verified-healthy startup window. Buffered data support electrochemically constrained trajectory matching, followed by calibrated state replay and residual-channel reactivation. Frozen cross-cycle healthy-residual calibration supports commissioning and remains in the voltage prediction after handover. Evaluation covers 216 additive-bias and 168 multiplicative-gain profiles over 24 A123 temperature--drive-cycle trajectories. Relative to Single-EKF, the original RBC-DEKF reduces mean SOC RMSE from 7.646 to 0.170 percentage points for additive bias and from 22.646 to 0.356 percentage points for gain faults. Separately, 18 nonzero-initial-error profiles on three representative 50\%-SOC-anchored segments give a full-record mean SOC RMSE of 1.747 percentage points, including startup, for the commissioned realization, versus 11.192 and 11.400 for the always-on RBC-DEKF and conventional joint EKF. Representative delayed bias, 10-s ramp, and gain tests show that the corrected SOC trajectory remains stable after fault introduction. The central contribution is electrochemically constrained temporal coordination of uncertain-state recovery and subsequent voltage-discrepancy accommodation.
\end{revisionblock}
\end{abstract}


\end{frontmatter}

\section{Introduction}

Lithium iron phosphate (LFP) has become one of the most widely deployed cathode materials in lithium-ion batteries for electric transportation and stationary energy-storage applications, owing to its intrinsic safety, long cycle life, thermal stability, and cost-effectiveness \cite{qi2024advances,Padhi1997,han2025micro}. Despite these advantages, reliable state estimation remains particularly challenging for LFP batteries \rev{\cite{liu2026physics,xiong2026hybrid}}. Lithium insertion and extraction proceed through a two-phase reaction between Li-rich LiFePO$_4$ and Li-poor FePO$_4$, producing an extended and nearly flat open-circuit-voltage (OCV) plateau over a broad state-of-charge (SOC) range \cite{Dreyer2010,zhang2025robust}. Within this plateau, the terminal voltage changes only weakly with SOC, substantially reducing voltage-based state observability \cite{ni2023rapid,zhang2025robust}. Consequently, a small voltage uncertainty can correspond to a large SOC ambiguity, making voltage-corrected SOC estimation intrinsically more difficult for LFP batteries than for chemistries with steeper OCV--SOC characteristics.

This weak observability places stringent requirements on the accuracy and reliability of voltage measurements used by battery management systems (BMSs) \cite{Plett2004,tian2023deep,chen2024estimating,zhang2025robust}. In long-term operation, however, voltage sensors may suffer from calibration drift, reference instability, component aging, and environmental influences \cite{zhang2022novel,wei2025advancements,zhang2025graph,wang2025impact}\rev{, with recent studies further emphasizing sensor-enabled monitoring under extreme environments and coupled thermal--electrical aging stresses \cite{xiong2025sensors,wang2026aging}}. These abnormalities commonly appear as additive offsets at the millivolt level or as multiplicative gain errors that rescale the measured voltage \cite{zhang2022voltage}. In an LFP battery, such apparently small sensing errors can be mapped onto disproportionately large corrections of the weakly observable electrochemical states, leading to severe SOC-estimation degradation or loss of convergence. Since SOC is a central variable for power limitation, energy management, balancing, and safety supervision, the resulting error can propagate through multiple BMS functions and compromise system-level reliability \cite{song2026unsupervised,yu2023challenges}. Voltage-sensor robustness is therefore not merely an instrumentation issue, but a material-dependent state-estimation problem for flat-plateau LFP batteries.

Battery fault-diagnosis methods are commonly divided into signal-processing-based, data-driven, and model-based approaches. Signal-processing methods extract abnormal temporal or spectral features directly from measurements and are attractive because of their relatively simple implementation \cite{zhang2023early,yuan2023early}. Their performance, however, can be sensitive to measurement noise, operating-condition variation, and the choice of diagnostic thresholds. Data-driven methods have demonstrated strong nonlinear representation capability for voltage-abnormality detection and battery fault classification \cite{zhao2017fault,li2020battery,li2025fault}, but generally require representative fault data and may exhibit limited generalization to previously unseen cells, temperatures, and load profiles \cite{zhao2024review,hu2020advanced,xiong2018review}. Recent realistic and model-constrained learning studies have improved fault diagnosis under stochastic and heterogeneous operating conditions \cite{zhang2023realistic,cao2025modelconstrained}; nevertheless, transferring laboratory-trained diagnosis models to practical battery systems remains difficult because of changing operating distributions, sensing limitations, and multi-scale fault manifestations \cite{zhao2024review}.

Model-based approaches provide an alternative route by linking faults to the inconsistency between measured and model-predicted battery responses \cite{xu2025recent,couto2018internal}\rev{; structural residual analysis has also been developed to isolate concurrent electro--thermal faults in battery systems \cite{wang2026structural}}. Most existing methods employ equivalent circuit models (ECMs), whose low computational cost and simple calibration are attractive for online BMS implementation. However, their residuals combine sensor abnormalities with process disturbances, parameter errors, and structural model mismatch, which complicates fault interpretation and threshold design \cite{xu2025recent}. This problem is particularly severe for LFP batteries: within the flat OCV region, a persistent voltage residual can be incorrectly explained by a large change in SOC even when the actual source is a sensor fault or modeling error.

Electrochemical models offer a more physically consistent basis for separating voltage formation from internal-state evolution. The pseudo-two-dimensional model and its reduced-order variants explicitly represent lithium transport, interfacial kinetics, and electrode thermodynamics, thereby connecting terminal voltage to physically meaningful internal states \cite{Fuller1994,haghverdi2025review}. Although the full-order model is generally too computationally demanding for recursive onboard estimation, reduced-order models such as the single-particle model preserve the dominant electrochemical dynamics at substantially lower computational cost \cite{guo2025control,guo2024systematic}. A direct comparison on graphite/LFP cells reported an average voltage-prediction error of 51.5 mV for an ECM and 19.3 mV for a physics-based model over temperatures of 10--40$^\circ$C and rates of C/3--2C, illustrating the value of electrochemical descriptions for LFP voltage prediction \cite{lagnoni2024critical}.

Residual correction provides an additional mechanism for addressing the discrepancy that remains even with a physics-based model. Physics-guided residual-learning frameworks have shown that an electrochemical-model prior can be combined with learned correction terms to improve LFP SOC estimation over broad temperature ranges and dynamic operating conditions \cite{GUO2026167,che2026mechanisticresidual}. These studies demonstrate that model residuals contain useful information for correcting state estimates. However, learned residual models generally require representative offline training data and primarily target model-induced prediction errors. They do not explicitly formulate voltage-sensor bias and gain abnormalities within the measurement model or recursively estimate a voltage-discrepancy state for online accommodation of both fault forms.

Several fault-tolerant strategies more directly related to voltage measurements have also been investigated. Sensor-topology-based reconstruction can tolerate isolated voltage-sensor faults by exploiting redundant pack measurements \cite{xia2016fault}, but its applicability depends on measurement redundancy and fault configuration. Adaptive filtering can improve robustness against small voltage perturbations \cite{li2016combination}, although such perturbations are often treated as generic uncertainty rather than explicit sensor faults. For LFP batteries, region-dependent methods have adjusted the correction strategy according to the local OCV slope or selected different observers for different OCV ranges \cite{xiong2023state,peng2025state}. Bias-compensated SOC and state-of-health estimation has also been developed by exploiting transitions between OCV regions \cite{yi2024bias}. These approaches improve estimation within flat-voltage regions, but typically require SOC-region identification, observer switching, or convergence across multiple charge--discharge intervals. In parallel, additive bias and multiplicative gain faults have been jointly represented through state augmentation in ECM-based estimators \cite{zhang2022voltage}. Their performance nevertheless remains strongly dependent on model fidelity and excitation, while the interaction between voltage-sensor abnormalities and the material-specific weak observability of LFP electrochemical states has not been explicitly addressed.

\begin{revisionblock}
Incorrect SOC initialization has been addressed through several complementary estimation strategies. Closed-loop UKF estimation has demonstrated self-correction from a large initial offset in an LFP cell \cite{he2013ukfinit}, while nonlinear observers have provided convergence analysis and recovery tests under a 20-percentage-point initial error \cite{xia2014nonlinearobserver}. Moving-horizon estimation exploits a finite measurement history and physical constraints to reduce sensitivity to mismatched initial conditions \cite{ren2020mhe,shen2021jointmhe}, and adaptive robust filters improve convergence when the initial SOC, model parameters, and noise statistics are simultaneously uncertain \cite{wang2022arukf}. These studies demonstrate the value of temporal excitation and constrained state evolution for recovering an uncertain SOC prior.

Staged adaptation has also been used in battery state--parameter co-estimation. Voltage-error dead zones can temporarily suspend physical-parameter updates when the innovation is unsuitable for reliable adaptation, including during large SOC-initialization errors \cite{guo2018deadzone}. Such methods primarily regulate physical quantities such as capacity, impedance, or OCV-related coefficients. In the present fault-tolerant setting, the auxiliary state instead represents a voltage-domain discrepancy that must remain available for subsequent voltage-abnormality accommodation. Without a dedicated startup allocation strategy, this rapidly adaptive residual state can absorb an initial electrochemical-state mismatch and prevent SOC recovery.
\end{revisionblock}

\rev{These studies highlight a central design challenge: on the flat LFP plateau, electrochemical-state error, model discrepancy, and sensor abnormality can produce similar voltage innovations. A fault-tolerant estimator must therefore coordinate state recovery and voltage-discrepancy accommodation without allowing a rapidly adaptive residual state to absorb uncertain initialization.}

\begin{revisionblock}
Motivated by these considerations, this work develops an electrochemically constrained, commissioning-enabled fault-tolerant SOC-estimation framework for flat-plateau LFP batteries. The main contributions are threefold:

\begin{enumerate}
    \item \textbf{Material-specific fault sensitivity.}
    The voltage-error sensitivity of LFP SOC estimation is quantified by linking the flat full-cell OCV response with the paired-electrode electrochemical description. This establishes the material-level motivation for explicitly accommodating systematic voltage discrepancy in flat-plateau state estimation.

    \item \textbf{Commissioning-enabled state--residual coordination.}
    A sequential residual-bias compensation dual extended Kalman filter (RBC-DEKF) is developed with separately configurable electrochemical-state and voltage-discrepancy recursions. Under uncertain initialization, residual adaptation is temporarily suspended while an electrochemically consistent paired-electrode state is corrected from a verified-healthy buffered startup trajectory, after which residual accommodation is reactivated.

    \item \textbf{Systematic evaluation and operating-boundary analysis.}
    Additive-bias and multiplicative-gain accommodation is evaluated over 24 temperature--drive-cycle trajectories, with separate representative tests of large initial-SOC errors and abnormalities introduced during operation. Comparisons with Single-EKF, open-loop propagation, and conventional joint augmentation distinguish fault accommodation from model propagation and state-recovery capability.
\end{enumerate}
\end{revisionblock}

\section{Electrochemical and material basis for state estimation}

\subsection{Classical SPM foundation}
The electrochemical backbone adopted in this work is derived from the classical SPM, which captures the essential solid-phase diffusion and interfacial-reaction dynamics in the negative and positive electrodes while remaining substantially simpler than full P2D formulations. In the classical SPM, solid-state lithium diffusion within representative spherical particles is governed by Fick's law. The terminal voltage is obtained from the electrode open-circuit potentials together with charge-transfer overpotentials and lumped ohmic loss. The original continuous-time SPM formulation is summarized in Table~\ref{tab:spm_original}.

\begin{table}[H]
\centering
\caption{Classical continuous-time single-particle model (SPM)}
\label{tab:spm_original}
\begin{threeparttable}
\small
\begin{tabular}{p{0.24\textwidth} p{0.70\textwidth}}
\toprule
\textbf{Component} & \textbf{Governing Equation} \\
\midrule

Solid-phase diffusion &
$\displaystyle
\frac{\partial c_{s,i}(r,t)}{\partial t}
=
\frac{1}{r^2}
\frac{\partial}{\partial r}
\left(
D_{s,i}r^2
\frac{\partial c_{s,i}(r,t)}{\partial r}
\right),
\qquad i\in\{n,p\}
$
\\[10pt]

Initial condition &
$\displaystyle
c_{s,i}(r,0)=c_{i,0}
$
\\[8pt]

Symmetry condition at particle center &
$\displaystyle
\left.
\frac{\partial c_{s,i}(r,t)}{\partial r}
\right|_{r=0}
=0
$
\\[8pt]

Flux boundary at particle surface &
$\displaystyle
\left.
\frac{\partial c_{s,i}(r,t)}{\partial r}
\right|_{r=R_{s,i}}
=
-\frac{j_i(t)}{D_{s,i}}
$
\\[10pt]

Surface concentration &
$\displaystyle
c_{ss,i}(t)
=
c_{s,i}(R_{s,i},t)
$
\\[8pt]

Normalized surface concentration &
$\displaystyle
\widetilde{c}_{ss,i}(t)
=
\frac{c_{ss,i}(t)}{c_{\max,i}}
$
\\[8pt]

Interfacial molar flux &
$\displaystyle
j_i(t)
=
s_i
\frac{I(t)}
{F a_i A_i L_i},
\qquad
s_n=+1,\quad s_p=-1
$
\\[10pt]

Exchange current density &
$\displaystyle
i_{0,i}(t)
=
F r_{\mathrm{eff},i}c_{\max,i}
\sqrt{
c_e
\widetilde{c}_{ss,i}(t)
\left[1-\widetilde{c}_{ss,i}(t)\right]
}
$
\\[10pt]

Charge-transfer overpotential &
$\displaystyle
\eta_i(t)
=
\frac{2RT}{F}
\sinh^{-1}
\left(
\frac{Fj_i(t)}
{2i_{0,i}(t)}
\right)
=
\frac{2RT}{F}
\sinh^{-1}
\left(
\frac{s_iI(t)}
{2a_iA_iL_i i_{0,i}(t)}
\right)
$
\\[12pt]

Terminal voltage &
$\displaystyle
V_{\mathrm{SPM}}(t)
=
OCP_p\!\left(\widetilde{c}_{ss,p}(t)\right)
-
OCP_n\!\left(\widetilde{c}_{ss,n}(t)\right)
+
\eta_p(t)-\eta_n(t)
-
R_{\mathrm{ini}}I(t)
$
\\

\bottomrule
\end{tabular}

\begin{tablenotes}[flushleft]
\footnotesize
\item
The applied current is defined as positive during discharge.
Accordingly, $s_n=+1$ and $s_p=-1$.
Here, $j_i$ denotes the interfacial molar flux
($\mathrm{mol\,m^{-2}\,s^{-1}}$), whereas $i_{0,i}$ denotes the
exchange current density ($\mathrm{A\,m^{-2}}$).
\end{tablenotes}

\end{threeparttable}
\end{table}

\subsection{Control-oriented parameter-grouped electrochemical model}

Although the classical SPM preserves the main electrochemical mechanisms, its direct use in recursive estimation is still inconvenient for real-time battery-management applications because the solid-phase diffusion dynamics are described by partial differential equations. To obtain a model suitable for online filtering, the diffusion dynamics are approximated by a parabolic concentration profile, and the original physical parameters are reformulated into grouped parameters that are more amenable to parameter identification and state estimation. This results in a reduced-order control-oriented parameter-grouped single-particle model with thermal effects (CPG-SPMT) \cite{guo2025control,GUO2026110075,guo2025comparative}, which retains the main electrochemical structure of the SPM while significantly improving computational efficiency. The resulting CPG-SPMT is summarized in Table~\ref{tab:cpg_spmt}.

The resulting reduced-order model can be written in a compact state-space form that is suitable for recursive estimation:
\begin{align}
\dot{\boldsymbol{\chi}}(t) &= \tilde{A}\,\boldsymbol{\chi}(t)+\tilde{B}\,u(t),
\qquad \boldsymbol{\chi}(0)=\boldsymbol{\chi}_0, \label{eq:cpg_spmt_state} \\
\boldsymbol{\psi}(t) &= \tilde{C}\,\boldsymbol{\chi}(t)+\tilde{D}\,u(t), \label{eq:cpg_spmt_output}
\end{align}
where $u(t)=I(t)$ is the applied current, $\boldsymbol{\chi}$ collects the reduced diffusion states of both electrodes, and $\boldsymbol{\psi}$ contains the normalized average and surface concentrations used in the nonlinear voltage calculation.

\begin{table}[H]
\centering
\caption{Control-oriented parameter-grouped single-particle model with thermal effects (CPG-SPMT)}
\label{tab:cpg_spmt}
\begin{threeparttable}
\small
\begin{tabular}{p{0.24\textwidth} p{0.70\textwidth}}
\toprule
\textbf{Item} & \textbf{Expression} \\
\midrule

State-space model &
$\displaystyle
\dot{\boldsymbol{\chi}}(t)=\tilde{A}\boldsymbol{\chi}(t)+\tilde{B}u(t),
\qquad
\boldsymbol{\psi}(t)=\tilde{C}\boldsymbol{\chi}(t)+\tilde{D}u(t)
$
\\[10pt]

State vector &
\rev{$\displaystyle
\boldsymbol{\chi}=
\begin{bmatrix}
\boldsymbol{\chi}_n^\top & \boldsymbol{\chi}_p^\top
\end{bmatrix}^\top,
\qquad
\boldsymbol{\chi}_i=
\begin{bmatrix}
\tilde{q}_{1,i} & \tilde{q}_{2,i}
\end{bmatrix}^\top,
\quad i\in\{p,n\}
$}
\\[12pt]

Output vector &
\rev{$\displaystyle
\boldsymbol{\psi}=
\begin{bmatrix}
\boldsymbol{\psi}_n^\top & \boldsymbol{\psi}_p^\top
\end{bmatrix}^\top,
\qquad
\boldsymbol{\psi}_i=
\begin{bmatrix}
\tilde{\overline{c}}_{s,i} & \tilde{c}_{ss,i}
\end{bmatrix}^\top
$}
\\[12pt]

Grouped system matrices &
$\displaystyle
\tilde{A}_i=
\begin{bmatrix}
0 & 0 \\
\frac{30}{\alpha_i} & -\frac{30}{\alpha_i}
\end{bmatrix},
\qquad
\tilde{B}_i=
\begin{bmatrix}
\frac{1}{b_i} \\
\frac{19}{7b_i}
\end{bmatrix},
\qquad
\tilde{C}_i=I_2,
\qquad
\tilde{D}_i=
\begin{bmatrix}
0 \\
\frac{\alpha_i}{105b_i}
\end{bmatrix}
$
\\[14pt]

Grouped transport parameters &
$\displaystyle
\alpha_i=\frac{R_{s,i}^2}{D_{s,i}},
\qquad
b_i=F A_i L_i \varepsilon_i c_{\max,i}
$
\\[12pt]

Grouped reaction overpotential &
$\displaystyle
\eta_i=
\frac{2RT}{F}
\sinh^{-1}
\left(
\frac{s_i I}
{6b_i d_i \sqrt{\tilde{c}_{ss,i}(1-\tilde{c}_{ss,i})}}
\right)
$
\\[14pt]

Grouped kinetic parameter &
$\displaystyle
d_i=
\frac{r_{\mathrm{eff},i}\sqrt{c_e}}
{F R_{s,i}}
$
\\[10pt]

Terminal voltage &
$\displaystyle
V=
OCP_p(\tilde{c}_{ss,p})
-
OCP_n(\tilde{c}_{ss,n})
+
\eta_p-\eta_n
-
R_{\mathrm{ini}}I
$
\\

\bottomrule
\end{tabular}
\begin{tablenotes}[flushleft]
\footnotesize
\item The current sign convention is identical to Table~\ref{tab:spm_original}: $s_n=+1$ and $s_p=-1$ for positive discharge current.
\end{tablenotes}
\end{threeparttable}
\end{table}

To account for thermal variation over a wide operating range, the key grouped transport and kinetic parameters are modeled as temperature-dependent using Arrhenius-type relations.
Specifically, the grouped diffusion parameters $\alpha_n$ and $\alpha_p$ vary with temperature as
\begin{align}
\alpha_n(T) &=
\frac{\alpha_{n,1}}
{\exp\!\left(
\frac{E_1}{R}\left(\frac{1}{T_{ref}}-\frac{1}{T}\right)
\right)}, \\
\alpha_p(T) &=
\frac{\alpha_{p,1}}
{\exp\!\left(
\frac{E_2}{R}\left(\frac{1}{T_{ref}}-\frac{1}{T}\right)
\right)},
\end{align}
the grouped reaction parameters $d_n$ and $d_p$ are expressed as
\begin{align}
d_n(T) &=
d_{n,1}\exp\!\left(
\frac{E_3}{R}\left(\frac{1}{T_{ref}}-\frac{1}{T}\right)
\right), \\
d_p(T) &=
d_{p,1}\exp\!\left(
\frac{E_4}{R}\left(\frac{1}{T_{ref}}-\frac{1}{T}\right)
\right),
\end{align}
and the lumped ohmic resistance is modeled by
\begin{equation}
R_{\mathrm{ini}}(T)=
\frac{R_{\mathrm{ini},1}}
{\exp\!\left(
\frac{E_5}{R}\left(\frac{1}{T_{ref}}-\frac{1}{T}\right)
\right)}.
\end{equation}
Here, $E_1$--$E_5$ denote the corresponding activation energies and $T_{ref}$ is the reference temperature.
\rev{These Arrhenius-type relations allow temperature dependence to be represented over the tested range.} By contrast, constants such as the universal gas constant $R$, Faraday's constant $F$, and electrolyte concentration $c_e$ are treated as known and fixed. The identified model parameters are listed in Table~\ref{tab:identified_parameters}. \rev{For reproducibility, \ref{app:parameter_identification} now gives the complete LFP-specific data matrix, 14-parameter bounds, bound-constrained multi-start refinement, and cross-condition selection criterion. The optimization workflow follows the batch-optimized local-to-global (BOLT) strategy~\cite{guo2026bolt}.}

\begin{table}[H]
\centering
\scriptsize
\setlength{\tabcolsep}{2pt}
\caption{Identified model parameters used in the CPG-SPMT}
\label{tab:identified_parameters}
\begin{threeparttable}
\begin{tabular}{p{0.13\textwidth} p{0.45\textwidth} p{0.11\textwidth} p{0.11\textwidth}}
\toprule
\textbf{Symbol} & \textbf{Description} & \textbf{Value} & \textbf{Unit} \\
\midrule

$\alpha_{n,1}$ & Reference negative-electrode diffusion grouped parameter at $25^\circ$C & 2696.83 & s \\
$\alpha_{p,1}$ & Reference positive-electrode diffusion grouped parameter at $25^\circ$C & 1352.43 & s \\
$b_n$ & Grouped negative-electrode capacity parameter & 10999.93 & C \\
$b_p$ & Grouped positive-electrode capacity parameter & 3982.26 & C \\
$d_{n,1}$ & Reference negative-electrode reaction grouped parameter at $25^\circ$C & 0.005475 & \makecell[l]{s$\cdot$mol$^{1/2}$\\m$^{-5/2}$} \\
$d_{p,1}$ & Reference positive-electrode reaction grouped parameter at $25^\circ$C & 0.000343 & \makecell[l]{s$\cdot$mol$^{1/2}$\\m$^{-5/2}$} \\
\rev{$s_{n,0}$} & \rev{Initial normalized negative-electrode stoichiometry} & 0.3829 & -- \\
\rev{$s_{p,0}$} & \rev{Initial normalized positive-electrode stoichiometry} & 0.000282 & -- \\
$E_1$ & Activation energy associated with $\alpha_n$ & 44042.22 & J/mol \\
$E_2$ & Activation energy associated with $\alpha_p$ & 52664.43 & J/mol \\
$E_3$ & Activation energy associated with $d_n$ & 133074.86 & J/mol \\
$E_4$ & Activation energy associated with $d_p$ & 14397.71 & J/mol \\
$E_5$ & Activation energy associated with $R_{\mathrm{ini}}$ & 4097.71 & J/mol \\
$R_{\mathrm{ini},1}$ & Reference ohmic resistance at $25^\circ$C & 0.1733 & $\Omega$ \\
$R$ & Universal gas constant & 8.314 & J/mol/K \\
$F$ & Faraday's constant & 96485 & C$\cdot$mol$^{-1}$ \\
$c_e$ & Electrolyte concentration & 1000 & mol$\cdot$m$^{-3}$ \\
\bottomrule
\end{tabular}
\end{threeparttable}
\end{table}

\begin{revisionblock}
The normalized average concentrations obtained from the reduced-order electrochemical model are finally converted into \emph{cell-equivalent electrode SOCs}. These variables are not the two raw lithiation fractions. Their orientations are selected so that both increase from 0 to 1 as the full-cell SOC increases, even though the underlying positive- and negative-electrode lithiation fractions evolve in opposite directions. Specifically,
\begin{align}
\mathrm{SOC}_p(t) &=
\frac{\tilde{\overline{c}}_{s,p}^{0\%}-\tilde{\overline{c}}_{s,p}(t)}
{\tilde{\overline{c}}_{s,p}^{0\%}-\tilde{\overline{c}}_{s,p}^{100\%}}, \label{eq:soc_positive}\\
\mathrm{SOC}_n(t) &=
\frac{\tilde{\overline{c}}_{s,n}(t)-\tilde{\overline{c}}_{s,n}^{0\%}}
{\tilde{\overline{c}}_{s,n}^{100\%}-\tilde{\overline{c}}_{s,n}^{0\%}}, \label{eq:soc_negative}
\end{align}
where $\tilde{\overline{c}}_{s,i}^{0\%}$ and $\tilde{\overline{c}}_{s,i}^{100\%}$ are the normalized electrode concentrations associated with the full-cell 0\% and 100\% SOC endpoints. For the positive electrode, $\tilde{\overline{c}}_{s,p}^{0\%}>\tilde{\overline{c}}_{s,p}^{100\%}$, which accounts explicitly for its reversed lithiation direction.

The overall cell SOC used in the estimator is then defined as
\begin{equation}
\mathrm{SOC}_{\mathrm{EKF}}(t)=\tfrac{1}{2}\big(\mathrm{SOC}_p(t)+\mathrm{SOC}_n(t)\big).
\label{eq:soc_cell}
\end{equation}

This cell-level SOC definition is used as the reported output of the estimator in the subsequent fault-tolerant estimation framework.

The initial electrochemical state is generated consistently from the cell-level prior $s_0\in[0,1]$ through
\begin{equation}
\hat{\mathbf{x}}_0=\boldsymbol{\phi}(s_0),\qquad
\begin{cases}
\tilde{\overline c}_{s,n}(0)=\tilde{\overline c}_{s,n}^{0\%}
+s_0\big(\tilde{\overline c}_{s,n}^{100\%}-\tilde{\overline c}_{s,n}^{0\%}\big),\\
\tilde{\overline c}_{s,p}(0)=\tilde{\overline c}_{s,p}^{0\%}
-s_0\big(\tilde{\overline c}_{s,p}^{0\%}-\tilde{\overline c}_{s,p}^{100\%}\big),
\end{cases}
\label{eq:initial_state_mapping}
\end{equation}
with a uniform initial solid concentration in each representative particle. This mapping is also used to impose the controlled initial-SOC perturbations in Section~\ref{subsec:initial_soc_recovery}.
\end{revisionblock}

\subsection{Material-specific voltage sensitivity and SOC ambiguity of LFP}
\label{subsec:lfp_voltage_sensitivity}

The flat voltage plateau of LFP is not only a characteristic of its two-phase electrochemical response; it also governs how strongly voltage uncertainty is translated into SOC uncertainty \cite{Dreyer2010,zhang2025robust,zhang2023novel}. To quantify this effect using the same battery platform considered in this work, the $25^\circ$C OCV characterization of the CALCE A1-007 LFP/graphite cell was analysed \cite{CALCE_A123_2025}. The approximately C/20 discharge branch was truncated at 2.5 V, and both low-rate branches were mapped to SOC using their measured capacities. After interpolation onto a common SOC grid, a branch-averaged quasi-equilibrium OCV was calculated as
\begin{equation}
U_{\mathrm{OCV}}(\mathrm{SOC})
\approx
\frac{1}{2}
\left[
V_{\mathrm{chg}}^{\mathrm{C/20}}(\mathrm{SOC})
+
V_{\mathrm{dis}}^{\mathrm{C/20}}(\mathrm{SOC})
\right].
\label{eq:branch_averaged_ocv}
\end{equation}
This averaging reduces the first-order influence of low-rate polarization and charge--discharge branch separation; it does not imply that hysteresis is absent. Before differentiation and inversion, the averaged curve was smoothed using a third-order Savitzky--Golay filter to suppress numerical noise while retaining the plateau structure.

\begin{figure}[H]
    \centering
    \includegraphics[width=\textwidth]{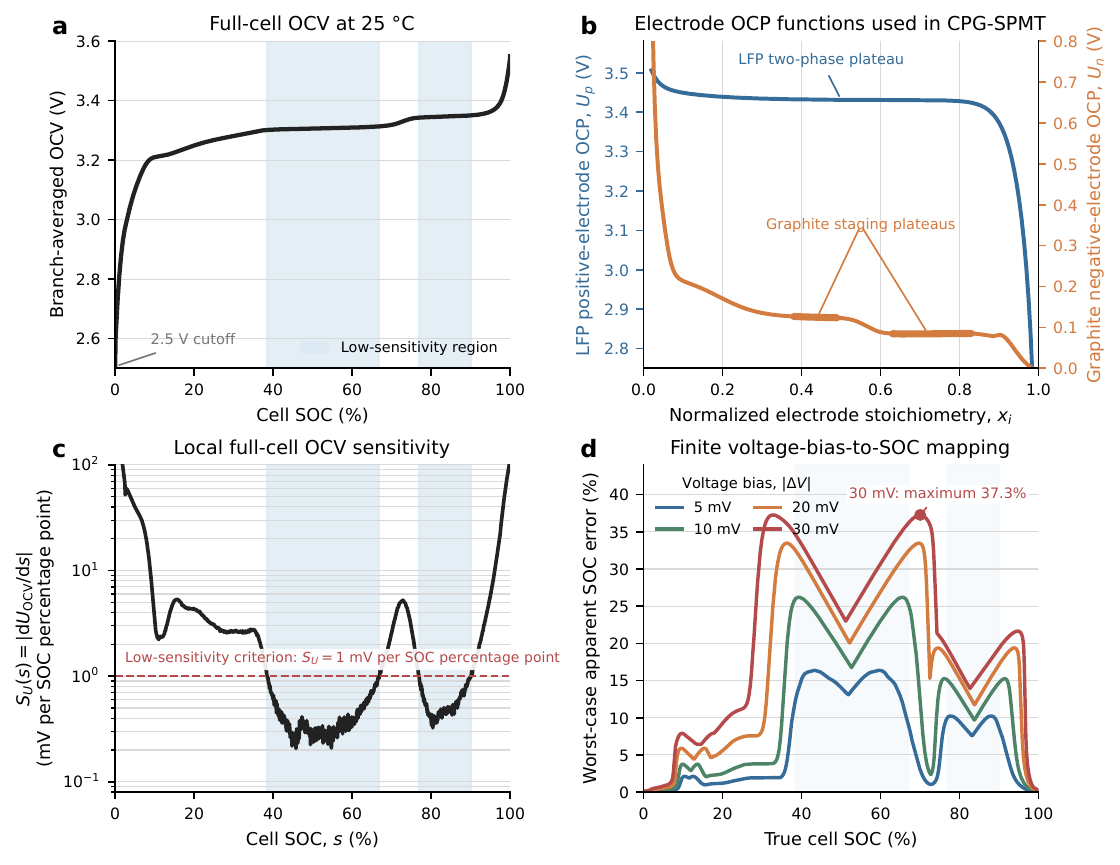}
    \caption{Material-specific voltage sensitivity of the A123 LFP/graphite cell at $25^\circ$C. (a) Branch-averaged full-cell OCV using a 2.5 V cutoff; shaded regions denote the low-sensitivity windows defined in panel (c). (b) Positive- and negative-electrode OCP functions used in the CPG-SPMT, with the LFP two-phase plateau and graphite staging plateaus indicated. (c) Local full-cell OCV sensitivity $S_U(s)=|\mathrm{d}U_{\mathrm{OCV}}/\mathrm{d}s|$; the dashed line defines the visualization criterion $S_U=1$ mV per SOC percentage point. (d) Worst-case apparent SOC error obtained by finite OCV inversion for the evaluated additive-bias magnitudes of 5--30 mV.}
    \label{fig:lfp_ocv_sensitivity}
\end{figure}

Figure~\ref{fig:lfp_ocv_sensitivity} connects the measured full-cell OCV response to the electrode-level OCP functions and the resulting voltage-to-SOC ambiguity. Figure~\ref{fig:lfp_ocv_sensitivity}(a) shows the branch-averaged OCV. The shaded windows are identical to those in panel (c), so that the full-cell plateau can be compared directly with the regions of low differential voltage sensitivity. Figure~\ref{fig:lfp_ocv_sensitivity}(b) shows the positive- and negative-electrode OCP functions used in the CPG-SPMT \cite{GUO2026110075}. The two curves are evaluated against their independent normalized stoichiometric coordinates and therefore should not be interpreted as a direct point-by-point decomposition of panel (a). The broad LFP positive-electrode plateau and the graphite staging plateaus jointly produce the nonuniform sensitivity of the full-cell voltage.

Let $s=100\,\mathrm{SOC}$ denote cell SOC in percentage points. The local full-cell voltage sensitivity is defined as
\begin{equation}
S_U(s)
=
\left|
\frac{\mathrm{d}U_{\mathrm{OCV}}}{\mathrm{d}s}
\right|,
\qquad
[S_U]=\mathrm{mV~per~SOC~percentage~point}.
\label{eq:local_ocv_slope}
\end{equation}
Under a local first-order linearization,
\begin{equation}
|\Delta s|
\approx
\frac{|\Delta V|}{S_U(s)}.
\label{eq:lfp_voltage_sensitivity}
\end{equation}
Figure~\ref{fig:lfp_ocv_sensitivity}(c) shows $S_U(s)$ over the full SOC range. For visualization, a low-sensitivity window is defined by $S_U<1$ mV per SOC percentage point. This criterion means that a 10 mV voltage perturbation corresponds locally to more than 10 SOC percentage points. It is used only to interpret the voltage response and is not used as a threshold by the RBC-DEKF. For the analysed OCV curve, the resulting windows are approximately 38.6--66.8\% SOC and 77.0--90.0\% SOC; the same windows are shaded in panels (a), (c), and (d).

To quantify finite perturbations without relying on the local approximation, the OCV-equivalent apparent SOC error is defined as
\begin{equation}
\varepsilon_{\mathrm{SOC}}(s,\Delta V)
=
\max_{\sigma\in\{-1,+1\}}
\left|
U_{\mathrm{OCV}}^{-1}
\!\left(U_{\mathrm{OCV}}(s)+\sigma\Delta V\right)-s
\right|,
\label{eq:finite_ocv_soc_error}
\end{equation}
where the perturbed voltage is restricted to the measured OCV range. Figure~\ref{fig:lfp_ocv_sensitivity}(d) reports the finite-inversion result for the additive-bias magnitudes evaluated later in this work: 5, 10, 20, and 30 mV. The maximum apparent SOC errors over the measured OCV range are approximately 16.4, 26.2, 33.5, and 37.3 pp, respectively. The strong amplification within the shaded windows provides a direct material-level explanation for the sensitivity of LFP SOC estimation to systematic voltage mismatch.

Equations~\eqref{eq:local_ocv_slope}--\eqref{eq:finite_ocv_soc_error} and Figure~\ref{fig:lfp_ocv_sensitivity} provide physical sensitivity interpretations rather than estimator equations. \rev{The branch-averaged quasi-equilibrium OCV reduces the first-order influence of charge--discharge branch separation in this static sensitivity analysis. The CPG-SPMT explicitly represents diffusion, kinetic polarization, ohmic loss, and temperature dependence, whereas dynamic voltage hysteresis is not explicitly modeled and therefore contributes to the remaining systematic model discrepancy.} The analysis nevertheless demonstrates why voltage mismatch is especially damaging in flat-voltage regions: forcing the electrochemical concentration states to explain such mismatch can create a large apparent SOC correction. \rev{The residual channel provides an alternative operational allocation for the dominant systematic discrepancy; it does not establish that the physical sources of that discrepancy are identifiable from voltage alone.}

\section{Electrochemically constrained fault-tolerant state estimation}

This section presents the proposed fault-tolerant state-estimation framework built upon the CPG-SPMT in Section~2.2 and the material-specific voltage sensitivity discussed in Section~2.3. The electrochemical states are constrained by lithium-transport dynamics in the positive and negative particles, whereas the dominant systematic voltage mismatch is represented by a separate scalar voltage-discrepancy state. \rev{This separation is an estimator design choice that allocates adaptation between two physically different roles; it neither adds an independent measurement nor removes the intrinsic weak observability of the LFP plateau.}

\begin{figure}[H]
    \centering
    \makebox[\textwidth][c]{\includegraphics[width=1.2\textwidth]{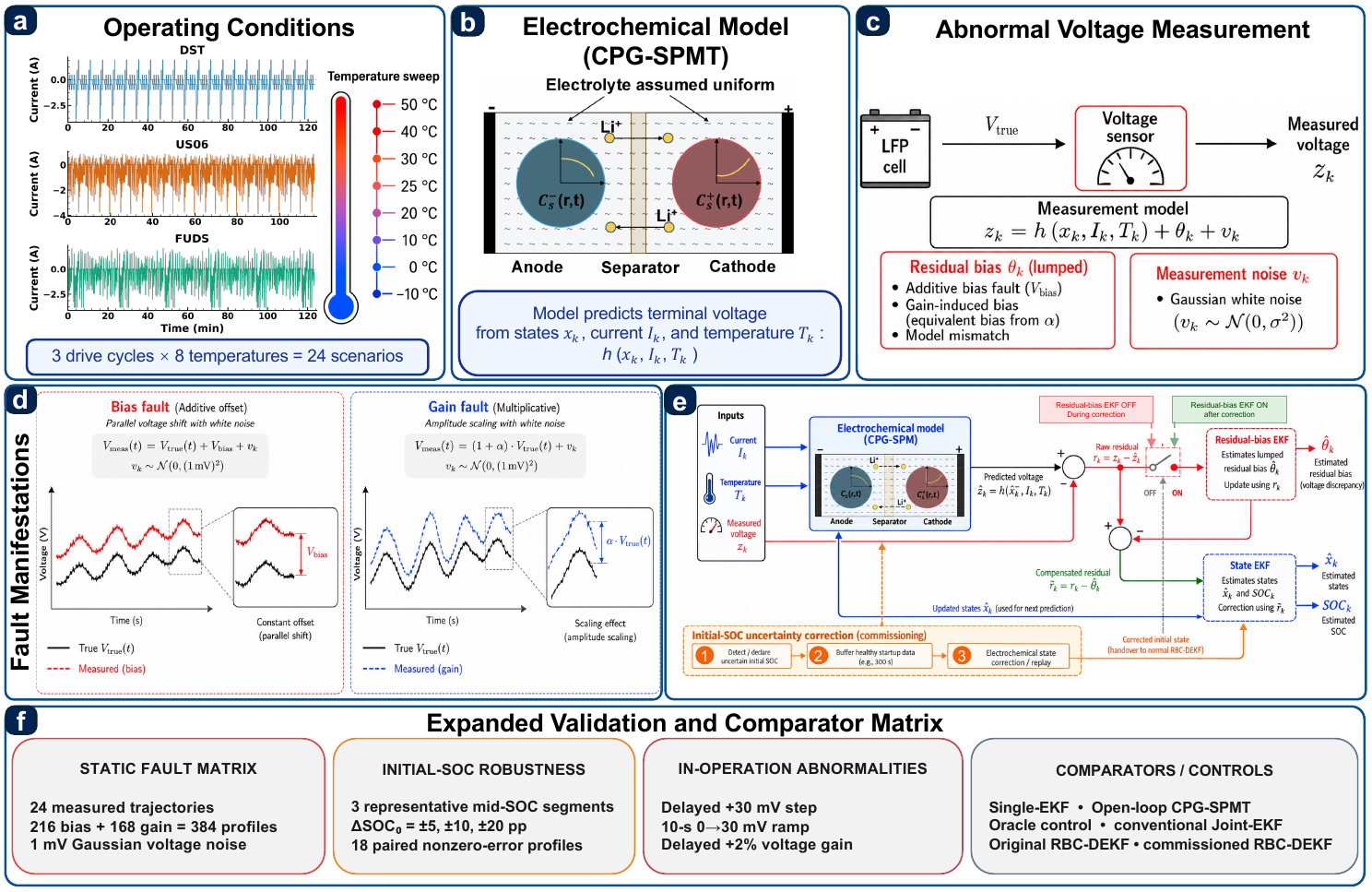}}
    \caption{\rev{Overview of the commissioning-enabled fault-tolerant framework. (a) Three drive cycles at eight temperatures. (b) Paired-electrode CPG-SPMT. (c) Voltage measurement with lumped discrepancy and noise. (d) Additive-bias and multiplicative-gain manifestations. (e) Functional state--residual coordination and uncertain-start commissioning; the state-first recursion is specified in Section~3.2 and commissioning is detailed in Figure~\ref{fig:commissioning_framework}. (f) Static faults, 50\%-SOC-anchored initialization tests, delayed abnormalities, and comparators. The 384 static profiles reuse 24 measured trajectories; the 18 nonzero-error initialization profiles and delayed-fault tests are separate.}}
    \label{fig:framework_overview}
\end{figure}

\rev{Figure~\ref{fig:framework_overview} summarizes the operating conditions, electrochemical model, voltage abnormalities, commissioning-enabled RBC-DEKF, and expanded validation matrix. The state-first sequential recursion is specified in Section~3.2, and the commissioning procedure and comparison estimators are defined in Sections~\ref{subsec:commissioning} and~\ref{subsec:baselines}, respectively.}

\subsection{Problem formulation and measurement model}
\label{subsec:state_ekf}

The objective of this work is to achieve accurate and fault-tolerant battery state estimation in the presence of abnormal voltage measurements (see Figure~\ref{fig:framework_overview}(c)).
The corresponding fault-tolerant estimation architecture is illustrated in Figure~\ref{fig:framework_overview}(e).
Let $z_k$ denote the measured terminal voltage at sample $k$, and let $h(\mathbf{x}_k,I_k,T_k)$ denote the nonlinear terminal-voltage output predicted by the electrochemical model introduced in Section~2.2, where $\mathbf{x}_k$ is the model state, $I_k$ is the applied current, and $T_k$ is the cell temperature.

In practice, the electrochemical model is only an approximation of the true battery behavior.
Therefore, the true terminal voltage can be written as
\begin{equation}
V_k^{\mathrm{true}} = h(\mathbf{x}_k,I_k,T_k) + \delta_k^{\mathrm{model}},
\label{eq:true_voltage}
\end{equation}
where $\delta_k^{\mathrm{model}}$ denotes the systematic voltage deviation caused by model mismatch, parameter uncertainty, temperature-dependent approximation error, and other unmodeled effects.
Under healthy sensing conditions, the measured voltage is then expressed as
\begin{equation}
z_k = V_k^{\mathrm{true}} + v_k
     = h(\mathbf{x}_k,I_k,T_k) + \delta_k^{\mathrm{model}} + v_k,
\label{eq:meas_normal}
\end{equation}
where $v_k$ denotes measurement noise.

In this study, two representative abnormal measurement modes are considered.
\rev{The first is an additive bias fault, represented as an additive voltage-domain offset:}
\begin{equation}
z_k = V_k^{\mathrm{true}} + V_{\mathrm{bias},k} + v_k
    = h(\mathbf{x}_k,I_k,T_k) + \delta_k^{\mathrm{model}} + V_{\mathrm{bias},k} + v_k.
\label{eq:meas_bias}
\end{equation}
The second is a multiplicative gain fault:
\begin{equation}
z_k = (1+\alpha)\,V_k^{\mathrm{true}} + v_k
    = (1+\alpha)\big(h(\mathbf{x}_k,I_k,T_k) + \delta_k^{\mathrm{model}}\big) + v_k,
\label{eq:meas_gain}
\end{equation}
where $\alpha$ is the gain-fault coefficient.
Equation~\eqref{eq:meas_gain} can be equivalently rewritten as
\begin{equation}
z_k = h(\mathbf{x}_k,I_k,T_k) + \delta_k^{\mathrm{model}}
      + \alpha V_k^{\mathrm{true}} + v_k,
\label{eq:meas_gain_equiv}
\end{equation}
which shows that a gain fault can be represented in the voltage-residual domain by the scalar, time-varying term $\alpha V_k^{\mathrm{true}}$.
\rev{Although the underlying sensor abnormality is multiplicative, its instantaneous voltage contribution is scalar. The residual state can therefore represent the combined discrepancy containing the gain-induced term without explicitly identifying $\alpha$. Compared with a constant additive offset, the adaptation demand increases when that combined residual becomes larger or varies more rapidly with terminal voltage.}

Equations~\eqref{eq:meas_bias}--\eqref{eq:meas_gain_equiv} also indicate that the systematic voltage residual generally contains two coupled sources: model-induced voltage mismatch $\delta_k^{\mathrm{model}}$ and sensor-induced measurement abnormality.
As a result, the compensation term estimated later in this work should be interpreted as a lumped voltage-discrepancy term rather than a pure sensor-bias estimate.

The discrete-time state equation of the control-oriented electrochemical model is written as
\begin{revisionblock}
\begin{equation}
\mathbf{x}_{k} = A_d(T_k)\mathbf{x}_{k-1} + B_d(T_k) I_k + \mathbf{w}_k,
\label{eq:state_discrete}
\end{equation}
\end{revisionblock}
\rev{Here $k$ indexes a processed data row: the implementation propagates the previous state using that row's $I_k,T_k$, then evaluates its voltage against $z_k$. The initial checkpoint precedes row $k=0$ and is denoted $\mathbf x_{-1}$ in this sampled recursion.} The matrices $A_d(\cdot)$ and $B_d(\cdot)$ are the temperature-dependent discrete-time system matrices obtained by zero-order hold discretization, and $\mathbf{w}_k$ is the process noise.
The state vector contains the reduced diffusion states of the positive and negative electrodes, from which the cell SOC is calculated as described in Section~2.2.
The above formulation defines the abnormal voltage measurement modes considered in this work, while the specific injected fault levels used for experimental evaluation are introduced in Section~\ref{subsec:fault_protocol}.

\begin{revisionblock}
Locally, the voltage innovation can be approximated as
\begin{equation}
\nu_k \simeq H_{x,k}\,\delta\mathbf{x}_k+\delta\theta_k+v_k.
\label{eq:local_state_residual_competition}
\end{equation}
Equation~\eqref{eq:local_state_residual_competition} makes the central ambiguity explicit: one scalar innovation cannot, in general, uniquely attribute a persistent discrepancy to electrochemical-state error, sensor abnormality, unmodeled hysteresis, or other model error. The following architectures impose different dynamic and covariance structures on that allocation, but do not create new measurement information.
\end{revisionblock}

\subsection{Residual-bias compensation dual extended Kalman filter (RBC-DEKF) framework}
\label{subsec:rbc_dekf}

To improve robustness against abnormal voltage measurements, an electrochemically constrained fault-tolerant estimation paradigm is developed based on the model and sensitivity structure introduced in Section~2.
The proposed framework starts from a conventional extended Kalman filter (EKF) for state estimation and augments it with an online residual-bias compensation mechanism.
In this way, the estimator preserves the physical interpretability of the model-based observer while accounting for the dominant systematic component of the voltage residual.

For the baseline Single-EKF state estimator, the discrete-time prediction model is written as
\begin{revisionblock}
\begin{align}
\hat{\mathbf{x}}_{k}^{-} &= A_d(T_k) \hat{\mathbf{x}}_{k-1} + B_d(T_k) I_k, \\
P_k^{-} &= A_d(T_k) P_{k-1} A_d^\top(T_k) + Q_x,
\end{align}
\end{revisionblock}
where $\hat{\mathbf{x}}_{k}^{-}$ and $P_k^{-}$ are the a priori state estimate and covariance matrix, respectively, and $Q_x$ denotes the state process noise covariance.

Without residual-bias compensation, the predicted terminal voltage is
\begin{equation}
\hat{z}_k = h(\hat{\mathbf{x}}_k^{-}, I_k, T_k),
\end{equation}
and the corresponding innovation is
\begin{equation}
e_k = z_k - \hat{z}_k.
\end{equation}
Using the linearized observation matrix
\begin{equation}
H_{x,k} = \left.\frac{\partial h}{\partial \mathbf{x}}\right|_{\hat{\mathbf{x}}_k^{-}},
\end{equation}
the Kalman gain is obtained as
\begin{equation}
K_k = P_k^{-} H_{x,k}^\top \left(H_{x,k} P_k^{-} H_{x,k}^\top + R_x\right)^{-1},
\end{equation}
where $R_x$ is the measurement noise covariance.
The posterior state update is then
\begin{align}
\hat{\mathbf{x}}_k &= \hat{\mathbf{x}}_k^{-} + K_k e_k, \\
P_k &= (I-K_k H_{x,k})P_k^{-}(I-K_k H_{x,k})^\top + K_k R_x K_k^\top.
\end{align}

Although this standard EKF can provide satisfactory SOC estimation under healthy sensing conditions, it becomes vulnerable when a systematic abnormal voltage deviation is present.
In such cases, the innovation term mixes state-related model mismatch and fault-induced measurement deviation, so the filter tends to absorb these systematic residuals into the electrochemical state estimate, which eventually degrades SOC estimation accuracy.

\rev{To accommodate this systematic discrepancy, a lumped voltage-discrepancy state} is introduced into the observation equation. Let $\theta_k$ denote the dominant systematic component of the voltage residual to be estimated online, which may originate from sensor faults, model mismatch, or other unmodeled effects. The augmented observation model is then written as
\begin{equation}
z_k = h(\mathbf{x}_k,I_k,T_k) + \theta_k + v_k,
\label{eq:augmented_measurement}
\end{equation}
where $v_k$ is the measurement noise. In this formulation, sensor fault and model mismatch are not explicitly separated, but are absorbed into the lumped voltage-discrepancy state.

Under this formulation, the electrochemical state observer remains responsible for tracking battery dynamics, while the voltage-discrepancy state captures the dominant systematic component of the voltage mismatch.

Based on this augmented formulation, the proposed framework proceeds in two sequential steps at each sampling instant. Rather than augmenting the electrochemical states and the voltage-discrepancy term into a single higher-dimensional EKF, the proposed method updates them sequentially in two coupled EKF loops. \rev{This realization provides the modularity required here: state and residual adaptation retain separate covariance recursions and can be enabled at different times. A conventional joint EKF is included as the principal augmented-state baseline in Section~\ref{subsec:baselines}.}

Because both $\mathbf{x}_k$ and $\theta_k$ influence the same scalar voltage measurement, their contributions are not independently identifiable from a single measurement without additional structure. \rev{The electrochemical dynamics and scalar random walk regularize this allocation operationally, while the covariance settings determine the relative adaptation bandwidth. They do not establish source identifiability.}

First, the electrochemical state is updated using the current voltage-discrepancy estimate. At time step $k$, the most recent available voltage-discrepancy estimate, namely $\hat{\theta}_{k-1}$, is used in the state-update loop.
The a priori voltage prediction becomes
\begin{equation}
\hat{z}_k^- = h(\hat{\mathbf{x}}_k^-, I_k, T_k) + \hat{\theta}_{k-1}.
\end{equation}

The Jacobian with respect to the electrochemical state remains
\begin{equation}
H_{x,k}=
\left.\frac{\partial h}{\partial \mathbf{x}}\right|_{\hat{\mathbf{x}}_k^-}.
\end{equation}

The state innovation, innovation covariance, and gain are given by
\begin{align}
e_{x,k} &= z_k - \hat{z}_k^-, \\
S_{x,k} &= H_{x,k} P_k^- H_{x,k}^\top + R_x, \\
K_{x,k} &= P_k^- H_{x,k}^\top S_{x,k}^{-1}.
\end{align}
The posterior state update is
\begin{align}
\hat{\mathbf{x}}_k &= \hat{\mathbf{x}}_k^- + K_{x,k} e_{x,k}, \\
P_k &= (I-K_{x,k}H_{x,k})P_k^-(I-K_{x,k}H_{x,k})^\top + K_{x,k}R_xK_{x,k}^\top.
\end{align}

Second, after the electrochemical state has been corrected, the voltage-discrepancy state is updated using the remaining voltage residual.
Its evolution is modeled as a random walk:
\begin{equation}
\theta_k = \theta_{k-1} + w_{\theta,k},
\label{eq:theta_random_walk}
\end{equation}
where $w_{\theta,k}\sim\mathcal{N}(0,Q_\theta)$.
\rev{The random-walk model provides a flexible low-order stochastic description of an unknown systematic voltage discrepancy, while $Q_\theta$ determines the available adaptation bandwidth.}
The prediction step for the voltage-discrepancy state is therefore
\begin{align}
\hat{\theta}_k^- &= \hat{\theta}_{k-1}, \\
P_{\theta,k}^- &= P_{\theta,k-1}+Q_\theta.
\end{align}
Using the updated state estimate, the \rev{compensated voltage prediction} is
\begin{equation}
\hat{z}_{\theta,k} = h(\hat{\mathbf{x}}_k,I_k,T_k)+\hat{\theta}_k^-.
\end{equation}
Because the voltage-discrepancy term enters the observation model additively, the corresponding Jacobian is simply
\begin{equation}
H_{\theta,k} = 1.
\end{equation}
The voltage-discrepancy innovation, innovation covariance, and gain are then
\begin{align}
e_{\theta,k} &= z_k-\hat{z}_{\theta,k}, \\
S_{\theta,k} &= P_{\theta,k}^{-}+R_\theta, \\
K_{\theta,k} &= P_{\theta,k}^{-} S_{\theta,k}^{-1},
\end{align}
and the posterior update is obtained as
\begin{align}
\hat{\theta}_k &= \hat{\theta}_k^- + K_{\theta,k} e_{\theta,k}, \\
P_{\theta,k} &= (1-K_{\theta,k})P_{\theta,k}^{-}(1-K_{\theta,k})^\top + K_{\theta,k}R_\theta K_{\theta,k}^\top.
\end{align}

\rev{The estimated voltage-discrepancy state provides an auxiliary signal reflecting systematic voltage mismatch. Its interpretation and the information needed for fault-source isolation are discussed in Section~\ref{subsec:residual_interpretation}.}

This sequential structure constitutes the fixed-architecture RBC-DEKF. Its essential feature is the explicit functional separation of battery-state correction and systematic voltage-residual compensation: the electrochemical state estimator follows internal battery dynamics, whereas the voltage-discrepancy estimator adapts to the dominant systematic voltage deviation. \rev{Whether a particular innovation component is assigned to $\mathbf{x}$ or $\theta$ remains tuning- and excitation-dependent; Section~\ref{subsec:commissioning} therefore adds a temporal constraint for uncertain initialization.}

\begin{revisionblock}
In all main experiments, the voltage-discrepancy state is initialized without a preloaded correction:
\begin{equation}
\hat{\theta}_0=0.
\label{eq:theta_init}
\end{equation}
The former first-residual warm start, $\hat\theta_0=z_0-h(\hat{\mathbf{x}}_0,I_0,T_0)$, is retained only as an initialization ablation in Section~\ref{subsec:overall_results}.
\end{revisionblock}
At each sampling instant, the proposed estimator follows the sequence:
\begin{enumerate}
    \item predict the electrochemical state and covariance;
    \item update the electrochemical state using the current voltage-discrepancy estimate;
    \item predict the voltage-discrepancy state;
    \item update the voltage-discrepancy state using the residual computed from the corrected state;
    \item calculate SOC from the updated electrochemical state.
\end{enumerate}
\rev{Compared with the conventional Single-EKF, the fixed-architecture RBC-DEKF introduces only one additional scalar variable and its associated covariance update, while preserving the same electrochemical state-space model.}
\rev{This adds only a scalar recursion and is amenable to online implementation.}
\rev{For additive sensor faults, the estimated voltage-discrepancy term accommodates the combined discrepancy containing the measurement offset. For multiplicative gain faults, it recursively represents a combined residual containing $\alpha V_k^{\mathrm{true}}$ and systematic model discrepancy. Because both enter as a scalar voltage-domain quantity, a one-dimensional compensation state can represent their combined instantaneous effect without explicitly estimating $\alpha$. The practical fit is governed by residual magnitude and temporal bandwidth and by the fixed covariance tuning of the residual loop.}

\subsection{Commissioning under uncertain initial SOC}
\label{subsec:commissioning}

\begin{figure}[htbp]
    \centering
    \includegraphics[width=0.98\textwidth]{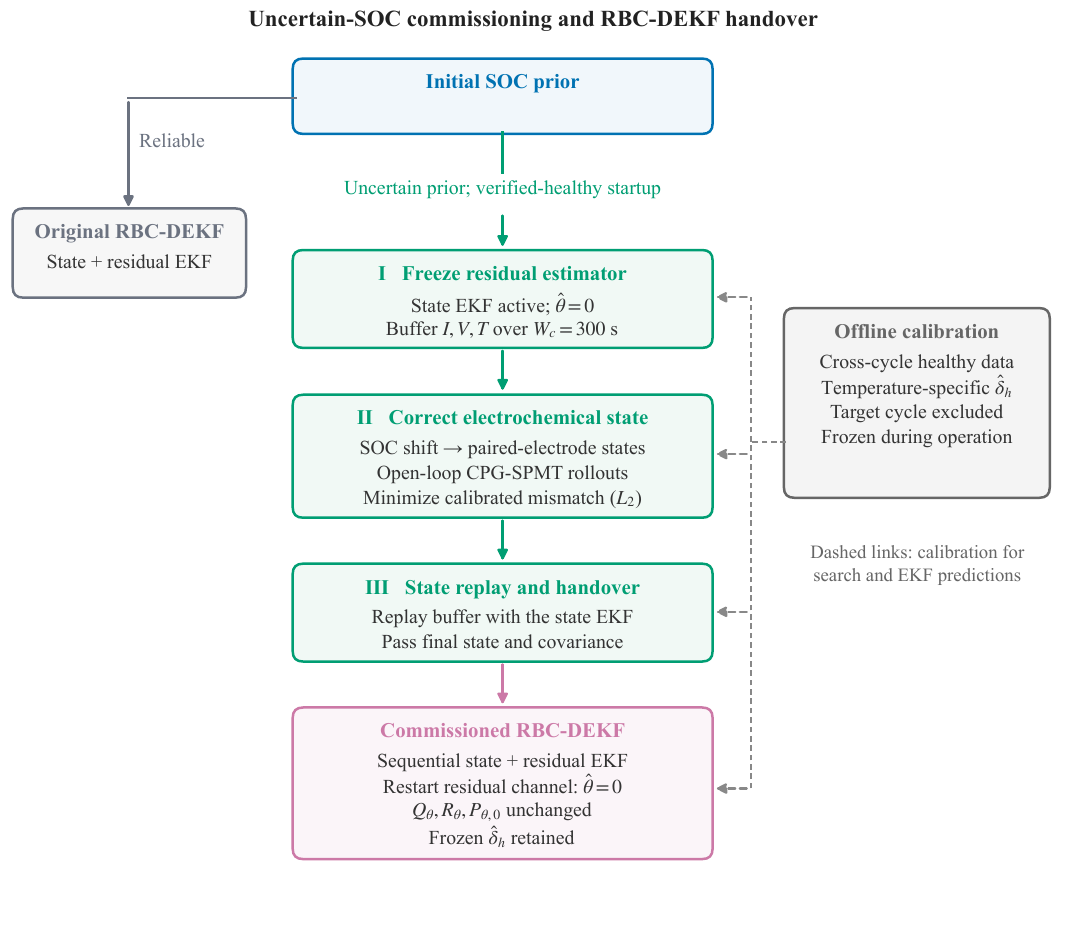}
    \caption{\rev{Commissioning and handover under uncertain initial SOC. The main path suspends the residual estimator, corrects the paired-electrode state using buffered healthy data, and replays the state EKF before activating sequential RBC-DEKF operation. The gray side module supplies frozen healthy-model calibration to the commissioned measurement predictions. A reliable initial prior enters the original RBC-DEKF directly. Residual-loop settings $Q_\theta$, $R_\theta$, and $P_{\theta,0}$ are unchanged.}}
    \label{fig:commissioning_framework}
\end{figure}

\begin{revisionblock}
When the initial SOC prior is reliable, the original RBC-DEKF is used directly. When the prior is declared uncertain and a healthy startup window is independently verified, commissioning temporarily suspends the voltage-discrepancy estimator, selects an electrochemically consistent initial state from the buffered measurements, and replays the corrected state before sequential residual accommodation begins. Figure~\ref{fig:commissioning_framework} summarizes these three online stages. A frozen healthy-model calibration supports the procedure; its construction and ablation are given in \ref{app:commissioning_ablation}.

\subsubsection{Commissioning principle}

During a window of $N_c=W_c/\Delta t$ samples ($W_c=300$ s and $\Delta t=1$ s here), residual prediction and correction are disabled:
\begin{equation}
\hat\theta_k=0,\qquad k=0,\ldots,N_c-1.
\label{eq:commissioning_suspend}
\end{equation}
The four-state EKF remains active, producing causal provisional SOC estimates and buffering current, temperature, and voltage. The BMS supplies both the uncertain-prior and healthy-window flags; a sensor abnormality already present at startup is outside this protocol.

A temperature-specific healthy-residual map $\hat\delta_{h,k}$ is fitted offline using the other two drive cycles at the same temperature, excluding the target cycle from training. It predicts repeatable healthy-model voltage discrepancy from SOC and causal current-history features and is frozen during operation. Reference SOC is used for offline fitting; candidate or estimated SOC is used online. The calibration remains in the commissioned voltage prediction after handover and is distinct from the adaptive voltage-discrepancy state.

\subsubsection{Electrochemically constrained state correction}

At window completion, candidate initial states are generated along the paired-electrode SOC direction $\mathbf m_s=\partial\boldsymbol\phi(s)/\partial s$ from Eq.~\eqref{eq:initial_state_mapping}. Let $\hat{\mathbf x}_{\mathrm{pr}}$ denote the checkpoint immediately before the first processed sample. Using the sample convention of Section~\ref{subsec:rbc_dekf}, a candidate is initialized as
\begin{equation}
\tilde{\mathbf x}_{-1}(\Delta s)=\hat{\mathbf x}_{\mathrm{pr}}
+\frac{\Delta s}{100}\mathbf m_s,
\label{eq:candidate_electrochemical_initial_state}
\end{equation}
where $\Delta s$ is in percentage points. The two entries of $\mathbf m_s$ for each electrode are equal, so the shift preserves the checkpoint's within-electrode reduced-state differences. Each candidate must generate a CPG-SPMT trajectory,
\begin{equation}
\tilde{\mathbf x}_{k}(\Delta s)=A_d(T_k)\tilde{\mathbf x}_{k-1}(\Delta s)
+B_d(T_k)I_k,\qquad k=0,\ldots,N_c-1.
\label{eq:candidate_electrochemical_trajectory}
\end{equation}
The provisional estimator and candidate search serve different roles: the former runs a causal state EKF, whereas the latter uses open-loop rollouts only to rank alternative initial electrochemical states.

Define $h_{c,k}(\mathbf x)=h(\mathbf x,I_k,T_k)+\hat\delta_{h,k}(s^{\mathrm{pct}}(\mathbf x))$, with $s^{\mathrm{pct}}$ the model-derived SOC in percentage units. The correction minimizes the calibrated voltage mismatch:
\begin{equation}
\Delta\hat s_W=\arg\min_{\Delta s\in\mathcal D_s}
\sum_{k=0}^{N_c-1}\left[z_k-h_{c,k}\!\left(\tilde{\mathbf x}_{k}(\Delta s)\right)\right]^2.
\label{eq:window_l2_objective}
\end{equation}
Here $\mathcal D_s$ is the 0.5-pp grid within $\lvert\Delta s\rvert\leq30$ pp and $0\leq\hat s_{\mathrm{pr}}+\Delta s/100\leq1$, where $\hat s_{\mathrm{pr}}$ is the checkpoint cell-SOC estimate as a fraction. The constant objective normalization used in code does not affect the minimizer. The retained search uses ordinary $L_2$ loss and applies its bounded minimizer directly. Both electrode OCPs and the dynamic polarization terms enter through $h$.

The calibration also enters the EKF gain through the measurement Jacobian:
\begin{equation}
H^c_{x,k}=H_{x,k}
+\frac{\partial\hat\delta_{h,k}}{\partial s^{\mathrm{pct}}}
\frac{\partial s^{\mathrm{pct}}}{\partial\mathbf x}.
\label{eq:calibrated_jacobian}
\end{equation}
The implemented piecewise-linear lookup slope is used in the provisional pass, corrected replay, and post-handover state updates; lookup and endpoint conventions are given in \ref{app:calibration_implementation}.

\subsubsection{Replay and handover}

The selected initial state is replayed through the buffered samples using the calibrated four-state EKF. Its terminal state and covariance become the handover condition. Earlier provisional outputs are retained; the replay endpoint is emitted when the window completes. From the next sample, the scalar residual EKF starts from zero with the original $Q_\theta$, $R_\theta$, and $P_{\theta,0}$ settings. The frozen calibration remains in both measurement predictions.

Retaining this calibration after handover keeps the measurement model consistent with state correction and replay, avoiding reintroduction of the repeatable healthy-model discrepancy at the commissioning boundary. The reactivated residual state accommodates deviations beyond the frozen prediction, including calibration error; these operational roles do not imply source separation.

The 300-s window is a fixed interior point in the observed 180--600-s stable-correction range on the tested trajectories (Section~\ref{subsec:window_sensitivity}). Hereafter, \emph{commissioned RBC-DEKF} denotes this complete calibrated realization. The contribution of this calibration is quantified by the ablation in \ref{app:commissioning_ablation}.
\end{revisionblock}

\subsection{Comparison estimators}
\label{subsec:baselines}

\begin{revisionblock}
Five estimator configurations are used across the evaluation, with the applicable subset stated for each experiment: (1) open-loop CPG-SPMT, (2) the four-state Single-EKF, (3) the original always-on RBC-DEKF, (4) a conventional five-state joint EKF, and (5) the commissioned RBC-DEKF. The open-loop model isolates electrochemical propagation without voltage feedback. The Single-EKF adds conventional voltage-based correction but contains no voltage-discrepancy state. The original RBC-DEKF is the sequential two-loop estimator of Section~\ref{subsec:rbc_dekf} and operates from the first sample. The joint EKF contains the same four electrochemical states and scalar residual state in one augmented update. The commissioned RBC-DEKF retains the sequential architecture but applies the uncertain-initialization procedure of Section~\ref{subsec:commissioning} before activating residual accommodation.

The static fault matrix compares the Single-EKF, original RBC-DEKF, and conventional joint EKF. The uncertain-initialization study additionally includes open-loop propagation and the commissioned RBC-DEKF. Specifically, the open-loop CPG-SPMT propagates
\begin{equation}
\hat{\mathbf{x}}_{k}=A_d(T_k)\hat{\mathbf{x}}_{k-1}+B_d(T_k)I_k
\label{eq:open_loop_baseline}
\end{equation}
without a voltage update. It distinguishes accurate physics propagation from genuine state-recovery capability. The Single-EKF is the four-state filter of Section~\ref{subsec:rbc_dekf} with $\theta\equiv0$. The original RBC-DEKF uses the state and residual recursions of Section~\ref{subsec:rbc_dekf} continuously from startup, without healthy-residual calibration or commissioning.

The conventional joint augmented EKF uses
\begin{equation}
\boldsymbol\xi_k=\begin{bmatrix}\mathbf{x}_k^\top&\theta_k\end{bmatrix}^{\!\top},\quad
F_k^J=\begin{bmatrix}A_d(T_k)&\mathbf{0}\\\mathbf{0}^\top&1\end{bmatrix},\quad
H_k^J=\begin{bmatrix}H_{x,k}&1\end{bmatrix},
\label{eq:joint_augmented_state}
\end{equation}
with $Q^J=\operatorname{blkdiag}(Q_x,Q_\theta)$ and $R_J=R_x=10^{-5}\,\mathrm{V}^2$. The joint estimator has one scalar measurement covariance; choosing $R_J=R_x$ matches the voltage weighting used by the electrochemical-state loop. Its single measurement update is
\begin{align}
S_k^J&=H_k^JP_k^{J,-}(H_k^J)^\top+R_J,\\
K_k^J&=P_k^{J,-}(H_k^J)^\top(S_k^J)^{-1},\\
\hat{\boldsymbol\xi}_k&=\hat{\boldsymbol\xi}_k^-+K_k^J
\big[z_k-h(\hat{\mathbf{x}}_k^-,I_k,T_k)-\hat\theta_k^-\big].
\label{eq:joint_ekf_update}
\end{align}
In particular, its state block satisfies
\begin{equation}
K_{x,k}^J=\frac{P_{xx,k}^-H_{x,k}^\top+P_{x\theta,k}^-}{S_k^J},
\label{eq:joint_state_gain}
\end{equation}
so the joint filter retains the full state--residual cross covariance. The principal augmented-state baseline is this conventional, always-on five-dimensional joint EKF initialized with $\hat\theta_0=0$.
\end{revisionblock}

\subsection{Sequential RBC-DEKF versus joint EKF: covariance coupling and commissioning capability}
\label{subsec:covariance_decoupling}

\begin{revisionblock}
The sequential RBC-DEKF and conventional joint EKF use the same four electrochemical states, scalar random-walk voltage-discrepancy state, and nonlinear voltage model. Their distinction lies in the allocation of covariance and measurement information. The RBC-DEKF stores separate state and residual design covariances and performs two ordered scalar-measurement updates. The joint EKF propagates a single $5\times5$ covariance, including the state--residual cross covariance, and corrects all five states with one common innovation.

For bookkeeping, the covariances explicitly stored by the sequential RBC-DEKF can be collected as
\begin{align}
\mathcal P_{k,\mathrm{stored}}^{D,-}
&=\operatorname{blkdiag}\!\left(P_k^-,P_{\theta,k}^-\right),
\label{eq:dual_stored_cov_prediction}\\
\mathcal P_{k,\mathrm{stored}}^{D}
&=\operatorname{blkdiag}\!\left(
J_{x,k}P_k^-J_{x,k}^{\top}+K_{x,k}R_xK_{x,k}^{\top},
(1-K_{\theta,k})^2P_{\theta,k}^-+K_{\theta,k}^2R_\theta
\right),
\label{eq:dual_stored_cov_update}
\end{align}
where $P_k^-=A_dP_{k-1}A_d^\top+Q_x$, $P_{\theta,k}^-=P_{\theta,k-1}+Q_\theta$, and $J_{x,k}=I_4-K_{x,k}H_{x,k}$. No state--residual cross-covariance block is propagated. Consequently,
\begin{equation}
K_{x,k}^{D}=\frac{P_k^-H_{x,k}^{\top}}
{H_{x,k}P_k^-H_{x,k}^{\top}+R_x},
\qquad
\left.\frac{\partial K_{x,k}^{D}}{\partial P_{\theta,k}^-}
\right|_{P_k^-,H_{x,k}}=0.
\label{eq:dual_covariance_decoupling}
\end{equation}
Equations~\eqref{eq:dual_stored_cov_prediction}--\eqref{eq:dual_stored_cov_update} describe the two design covariances maintained by the algorithm; they do not assert statistical independence of the physical estimation errors. The numerical estimates remain coupled because the state innovation uses the preceding residual estimate and the residual innovation is evaluated after the state correction. Because the same voltage sample is used in two separately tuned sequential updates, RBC-DEKF is not a covariance-equivalent factorization of the conventional joint EKF.

The independence of $K_{x,k}^{D}$ from $P_{\theta,k}^{-}$ is the operational decoupling used during commissioning. The scalar residual prediction and update can be suspended while retaining the four-state EKF prediction/update structure and Joseph covariance recursion. Electrochemical-state recovery can therefore proceed without residual-state uncertainty entering the state-loop gain directly.

By contrast, partition the conventional joint-EKF covariance as
\begin{equation}
P_k^J=\begin{bmatrix}
P_{xx,k}^J & P_{x\theta,k}^J\\
P_{\theta x,k}^J & P_{\theta\theta,k}^J
\end{bmatrix}.
\label{eq:joint_cov_partition}
\end{equation}
With $A_k=A_d(T_k)$, its prediction propagates every block:
\begin{equation}
P_k^{J,-}=\begin{bmatrix}
A_kP_{xx,k-1}^JA_k^{\top}+Q_x & A_kP_{x\theta,k-1}^J\\
P_{\theta x,k-1}^JA_k^{\top} & P_{\theta\theta,k-1}^J+Q_\theta
\end{bmatrix}.
\label{eq:joint_cov_prediction_expanded}
\end{equation}
The scalar innovation covariance and gain blocks therefore become
\begin{align}
S_k^J={}&H_{x,k}P_{xx,k}^{J,-}H_{x,k}^{\top}
+H_{x,k}P_{x\theta,k}^{J,-}+P_{\theta x,k}^{J,-}H_{x,k}^{\top}
+P_{\theta\theta,k}^{J,-}+R_J,
\label{eq:joint_innovation_cov_expanded}\\
\begin{bmatrix}K_{x,k}^J\\K_{\theta,k}^J\end{bmatrix}
={}&\frac{1}{S_k^J}\begin{bmatrix}
P_{xx,k}^{J,-}H_{x,k}^{\top}+P_{x\theta,k}^{J,-}\\
P_{\theta x,k}^{J,-}H_{x,k}^{\top}+P_{\theta\theta,k}^{J,-}
\end{bmatrix}.
\label{eq:joint_gain_blocks}
\end{align}
Its full posterior covariance is updated in Joseph form,
\begin{equation}
P_k^J=J_k^JP_k^{J,-}(J_k^J)^{\top}+K_k^JR_J(K_k^J)^{\top},
\qquad J_k^J=I_5-K_k^JH_k^J.
\label{eq:joint_cov_joseph}
\end{equation}
Even if $P_{x\theta,0}^J=0$, the common voltage update generally creates a nonzero cross block; for a block-diagonal prior,
\begin{equation}
P_{x\theta,k}^{J}=-P_{xx,k}^{J,-}H_{x,k}^{\top}(S_k^J)^{-1}P_{\theta\theta,k}^{J,-}.
\label{eq:joint_cross_cov_generation}
\end{equation}
Thus, the joint EKF supplies a coherent first-order covariance for the complete augmented state, but both $P_{\theta\theta,k}^{J,-}$ and $P_{x\theta,k}^{J,-}$ participate in its common innovation covariance and state gain. A staged joint implementation is possible, but it requires an explicit mode-dependent constraint or masked augmented update together with a consistent rule for the cross-covariance block. The sequential RBC-DEKF exposes the state and residual recursions as independently controllable modules. This functional and temporal decoupling enables Section~\ref{subsec:commissioning}: the electrochemical state is corrected first under the verified-healthy constraint, and the voltage-discrepancy loop is activated after handover.
\end{revisionblock}

\begin{table}[H]
\centering
\color{revisionyellow}
\caption{Covariance treatment in the sequential RBC-DEKF and conventional five-dimensional joint EKF.}
\label{tab:dual_joint_covariance}
\small
\setlength{\tabcolsep}{4pt}
\begin{tabular}{>{\raggedright\arraybackslash}p{0.19\textwidth}>{\raggedright\arraybackslash}p{0.35\textwidth}>{\raggedright\arraybackslash}p{0.35\textwidth}}
\toprule
Property & Sequential RBC-DEKF & Joint augmented EKF \\
\midrule
Stored covariance & $P_k\in\mathbb R^{4\times4}$ and scalar $P_{\theta,k}$ & Full $P_k^J\in\mathbb R^{5\times5}$ \\
State--residual cross block & Not propagated & $P_{x\theta,k}^J$ is propagated and measurement-updated \\
State-loop gain & Depends on $P_k^-$, $H_{x,k}$, and $R_x$ & Also depends on $P_{\theta\theta,k}^{J,-}$ and $P_{x\theta,k}^{J,-}$ \\
Voltage weighting & Separate $R_x$ and $R_\theta$ in two sequential updates & One $R_J$ in the augmented update \\
Commissioning action & Skip the scalar residual recursion; retain the four-state covariance update & A comparable suspension requires an explicit mode-dependent constraint or masking, with consistent treatment of the state--residual cross-covariance block \\
Interpretation & Functional/temporal modularity; not statistical independence & Covariance-consistent joint augmentation; not source identifiability \\
\bottomrule
\end{tabular}
\end{table}

\subsection{Filter settings and implementation details}

Separate covariance settings are assigned to the electrochemical state estimator and the \rev{voltage-discrepancy loop}.
For the state estimator, the process noise covariance is set to
\begin{equation}
Q_x = 10^{-18} I_4,
\end{equation}
the measurement noise covariance is
\begin{equation}
R_x = 10^{-5},
\end{equation}
and the initial state covariance is
\begin{equation}
P_{x,0}=10^{-6}I_4.
\end{equation}
\rev{The very small $Q_x$ makes the recursive state estimate rely primarily on model propagation, initialization, and the limited voltage correction available on the LFP plateau. It is reported as a design setting rather than interpreted as a measured physical process-noise variance.}

For the \rev{voltage-discrepancy loop}, the process noise, measurement noise, and initial covariance are chosen as
\begin{align}
Q_\theta &= 5\times 10^{-3},\\
R_\theta &= 3\times 10^{-4},\\
P_{\theta,0} &= 3\times 10^{3}.
\end{align}
These covariance values were selected empirically and kept fixed for all operating conditions and fault levels. The large initial covariance enables rapid early-stage adaptation, while $Q_\theta$ provides sufficient responsiveness to systematic voltage deviations and $R_\theta$ regularizes the update against measurement noise. The voltage-related covariance quantities $R_x$, $Q_\theta$, $R_\theta$, and $P_{\theta,0}$ are expressed in $\mathrm{V}^2$. These quantities are filter-design covariances rather than direct estimates of the injected sensor-noise variance. In particular, $R_x$ represents the effective voltage-output uncertainty used by the state filter, including measurement noise, local linearization error, and residual model mismatch. \rev{For the uncertain-initialization study, the prior covariance is enlarged only along the SOC-aligned state direction as $P_{x,0}^{\mathrm{unc}}=10^{-6}I_4+\sigma_{s,0}^2\mathbf d_s\mathbf d_s^\top$, where $\mathbf d_s=\left.\partial\boldsymbol\phi(s)/\partial s\right|_{s=\hat s_0}$ and $\sigma_{s,0}=0.10$; all reported $Q/R/P_\theta$ settings remain unchanged.}

In all simulations, the sampling interval is $\Delta t=1$ s.
\rev{The Jacobian of the nonlinear voltage model with respect to the four electrochemical states is evaluated by forward finite differences with a state perturbation of $10^{-6}$, whereas the Jacobian with respect to the additive voltage-discrepancy state is exactly one.} The covariance matrices are updated using the Joseph form for improved numerical robustness.
The initial electrochemical state is constructed from the initial SOC estimate by mapping the cell-level SOC to the average concentrations of both electrodes and setting each initial surface concentration equal to the corresponding average concentration, which represents a uniform initial concentration profile within the solid particles.

\subsection{Fault injection protocol}
\label{subsec:fault_protocol}

Based on the abnormal measurement modes defined in Section~3.1, additive bias faults and multiplicative gain faults are injected independently into the measured terminal-voltage channel for evaluation, while the underlying current, temperature, and reference SOC trajectories remain unchanged. All fault tests are performed on top of white Gaussian voltage noise with zero mean and a standard deviation of 1 mV.

For additive bias faults, constant offset levels are imposed as
\begin{equation}
V_{\mathrm{bias},k} = b_f,
\qquad
b_f \in \{0,\pm5,\pm10,\pm20,\pm30\}\,\mathrm{mV}.
\end{equation}

For multiplicative gain faults, the imposed gain-fault levels are
\begin{equation}
\alpha = \alpha_f,
\qquad
\alpha_f \in \{0,\pm1\%,\pm2\%,\pm3\%\}.
\end{equation}

These injected abnormalities are evaluated under the operating scenarios described in Section~4.1.

\begin{revisionblock}
The original matrices above apply a constant abnormality from the first evaluated sample. To examine faults introduced during operation, three additional profiles begin at the first sample after the temporal midpoint, $k_f=\lceil N/2\rceil$, of a trajectory:
\begin{align}
b_k^{\mathrm{step}}&=\begin{cases}0,&k<k_f,\\30~\mathrm{mV},&k\geq k_f,\end{cases}\\
b_k^{\mathrm{ramp}}&=\begin{cases}0,&k<k_f,\\
30~\mathrm{mV}\min\!\left(\dfrac{k-k_f}{10},1\right),&k\geq k_f,\end{cases}\\
\alpha_k^{\mathrm{gain}}&=\begin{cases}0,&k<k_f,\\0.02,&k\geq k_f.\end{cases}
\label{eq:delayed_fault_profiles}
\end{align}
The ramp reaches $+30$ mV at $k_f+10$ because $\Delta t=1$ s. These tests evaluate accommodation rather than automatic detection. In the proposed commissioned RBC-DEKF, the residual loop is enabled immediately after the verified-healthy 300 s commissioning interval and remains active thereafter; the estimator is not supplied with $k_f$. The original RBC-DEKF and conventional joint EKF are active from startup and likewise receive no onset input. The time-domain comparisons use the $25^\circ$C--DST 50\%-SOC-anchored trajectory with a $+10$-percentage-point initial error, matching the initialization used in Section~\ref{subsec:delayed_faults}.
\end{revisionblock}

\section{Results}
\label{sec:results}

\subsection{Experimental settings and evaluation scenarios}
\label{subsec:exp_settings}

The proposed framework was evaluated using 24 measured trajectories from the A123 LFP/graphite cell A1-007 in the open-source CALCE dataset~\cite{CALCE_A123_2025}. The dataset provides synchronized current, voltage, and temperature measurements over a wide range of dynamic and thermal operating conditions.

The evaluation covered eight temperatures, namely $-10$, $0$, $10$, $20$, $25$, $30$, $40$, and $50^\circ\mathrm{C}$, together with three representative dynamic drive cycles, namely Dynamic Stress Test (DST), US06 Supplemental Federal Test Procedure (US06), and Federal Urban Driving Schedule (FUDS), resulting in 24 temperature--cycle trajectories. The complete settings are listed in Table~\ref{tab:exp_settings}, while Figure~\ref{fig:framework_overview}(f) provides a graphical summary. \rev{All 24 trajectories participate in the model-identification/candidate-selection matrix described in \ref{app:parameter_identification}; their reuse here therefore demonstrates broad in-matrix operating-condition coverage, not strict out-of-sample model generalization. The condition-wise model-fit audit in Table~\ref{tab:cpg_model_fit} gives macro-mean terminal-voltage MAE and RMSE values of 21.796 and 33.069 mV, respectively. The remaining healthy model residual is consequential and is not treated as a sensor fault.}

Based on the fault-injection settings defined in Section~\ref{subsec:fault_protocol}, additive bias faults included 9 levels in total ($0,\pm5,\pm10,\pm20,\pm30$ mV), yielding $24 \times 9 = 216$ additive-bias condition--fault profiles per applicable estimator. Multiplicative gain faults included 7 levels in total ($0,\pm1\%,\pm2\%,\pm3\%$), yielding $24 \times 7 = 168$ gain-fault profiles per applicable estimator. The no-fault condition was included separately in both fault families; accordingly, \rev{384 denotes condition--fault profiles constructed from 24 measured trajectories for each estimator included in that comparison, not 384 independent experiments or cells.}

\begin{revisionblock}
All initialization experiments use segments anchored at 50\% SOC. Each segment begins at the first sample for which the coulomb-counted reference SOC is less than or equal to 50\%; because the records are sampled discretely, the actual first-sample SOC can be slightly below 50\%. Hereafter, these records are termed \emph{50\%-SOC-anchored segments}.

Three additional studies are reported separately from the 384-profile static fault matrix because their units of analysis differ. The initial-SOC study uses three representative 50\%-SOC-anchored segments and seven initial perturbations ($0,\pm5,\pm10,\pm20$ percentage points), giving 21 paired trajectory--initialization profiles evaluated by four methods. Its primary nonzero-error metric uses the 18 profiles corresponding to the six nonzero perturbations, while the three zero-error controls are reported separately. The commissioning-window study evaluates five fixed horizons on the same 21 profiles, giving 105 commissioned-estimator profiles. Finally, the operational-onset study uses the $25^\circ$C--DST segment with a $+10$-percentage-point initial error and compares three methods under three delayed abnormalities.
\end{revisionblock}

For all estimator fault-injection and initialization evaluations, zero-mean Gaussian white noise with a standard deviation of 1 mV was added to the voltage measurement. \rev{The static fault matrix compares the Single-EKF, the original always-on RBC-DEKF, and the conventional joint EKF. The commissioned RBC-DEKF is evaluated separately in the uncertain-initialization studies of Sections~\ref{subsec:initial_soc_recovery}--\ref{subsec:delayed_faults}, which also include open-loop propagation as a state-propagation reference. For the filter-based estimates, the reported voltage RMSE compares the posterior fitted voltage with the same faulty measurement used in the update. The fitted voltage is $h(\hat{\mathbf{x}}_k)$ for the Single-EKF, $h(\hat{\mathbf{x}}_k)+\hat\theta_k$ for the original RBC-DEKF or conventional joint EKF, and $h(\hat{\mathbf{x}}_k)+\hat\delta_{h,k}+\hat\theta_k$ for the commissioned branch, with $\hat\theta_k=0$ while its residual channel is suspended. It is therefore a posterior faulty-measurement fit error, not a clean-voltage recovery metric.} The results are presented first through aggregated statistics over the full operating matrix, followed by representative time-domain cases, cross-condition coverage, and the initialization and fault-onset studies. For a reference sequence $y_k$ and its estimate $\hat{y}_k$, the root-mean-square error is defined as
\begin{equation}
\mathrm{RMSE}(y)=\sqrt{\frac{1}{N}\sum_{k=1}^{N}\left(\hat{y}_k-y_k\right)^2}.
\label{eq:rmse_metric}
\end{equation}
\rev{SOC RMSE and SOC error are reported in percentage points (pp), whereas posterior faulty-measurement fit RMSE is reported in mV.} For each matched operating condition, the relative improvement is computed as
\begin{equation}
\mathcal{I}=100\left(1-\frac{\mathrm{RMSE}_{\mathrm{RBC-DEKF}}}{\mathrm{RMSE}_{\mathrm{Single-EKF}}}\right)\%.
\label{eq:improvement_metric}
\end{equation}

\begin{table}[H]
\centering
\caption{Experimental settings and validation matrix. The static fault matrix and the additional initialization, window-sensitivity, and operational-onset studies use different analysis units and are therefore reported separately rather than summed.}
\label{tab:exp_settings}
\renewcommand{\arraystretch}{0.94}
\setlength{\tabcolsep}{3pt}
\scriptsize
\resizebox{\textwidth}{!}{%
\begin{tabular}{>{\raggedright\arraybackslash}p{0.31\textwidth} >{\raggedright\arraybackslash}p{0.61\textwidth}}
\toprule
\textbf{Item} & \textbf{Setting} \\
\midrule
Battery dataset & Open-source CALCE A123 LFP/graphite cell A1-007 \\
Cell chemistry & LFP cathode / graphite anode \\
Temperatures & $-10,\,0,\,10,\,20,\,25,\,30,\,40,\,50\,^\circ\mathrm{C}$ \\
Drive cycles & DST, US06, FUDS \\
Number of measured operating trajectories & $8 \times 3 = 24$ \\
\rev{Parameter identification matrix} & \rev{All 24 temperature--cycle trajectories (\ref{app:parameter_identification})} \\
Additive bias faults & $0,\pm5,\pm10,\pm20,\pm30$ mV \\
Number of bias levels & 9 \\
Additive-bias test cases & $24 \times 9 = 216$ \\
Multiplicative gain faults & $0,\pm1\%,\pm2\%,\pm3\%$ \\
Number of gain levels & 7 \\
Gain-fault test cases & $24 \times 7 = 168$ \\
Voltage noise & Gaussian white noise, $\sigma = 1$ mV \\
Sampling interval & 1 s \\
Static-matrix SOC domain & Record start through the first reference-SOC $\leq10\%$ crossing (inclusive), or available record end \\
Static-matrix methods & Single-EKF, original always-on RBC-DEKF ($\hat\theta_0=0$), and conventional five-state joint EKF \\
Initial-SOC comparison methods & Open-loop CPG-SPMT, original RBC-DEKF, conventional joint EKF, and commissioned RBC-DEKF \\
Initial-SOC recovery study & Three 50\%-SOC-anchored segments $\times$ seven perturbations = 21 paired profiles; primary metric: 18 nonzero profiles \\
Commissioning-window sensitivity & Five horizons ($W_c=60$--600 s) on the same 21 profiles = 105 commissioned-estimator profiles \\
Operational-onset study & One $25^\circ$C--DST, $+10$-pp case; step, 10-s ramp, and $+2\%$ gain, each evaluated by three methods \\
Main metrics & SOC RMSE; posterior faulty-measurement fit RMSE; improvement (\%) \\
Total static fault-matrix size & 384 profiles per applicable estimator; excludes the additional studies above \\
\bottomrule
\end{tabular}
}
\end{table}

\begin{revisionblock}
To characterize SOC dependence, Figure~\ref{fig:soc_residual_by_soc} bins $r_{\mathrm{model}}=V_{\mathrm{meas}}-V_{\mathrm{CPG-SPMT}}$ by coulomb-counted reference SOC over all 24 temperature--cycle records and reports trajectory-equal signed and RMS statistics. Across all full-record samples with coulomb-counted reference SOC $\geq10\%$, the condition-macro raw-model RMSE and MAE are 25.210 and 19.125 mV, respectively. These threshold-filtered values complement rather than replace the 33.069 and 21.796 mV complete-record values in Table~\ref{tab:cpg_model_fit}; they are not presented as a reconstruction of the static estimator's first-crossing truncation. The marked increase in the $0$--$10\%$ bin is interpreted cautiously because only 20 of the 24 trajectories reach that bin; the figure explicitly reports the number of contributing trajectories for every SOC interval.
\end{revisionblock}

\begin{table}[H]
\centering
\color{revisionyellow}
\caption{Condition-wise CPG-SPMT terminal-voltage model-fit errors over the 24 temperature--cycle trajectories. RMSE and MAE are in mV. Because every trajectory participates in parameter refinement and/or candidate selection, these values summarize in-matrix fit rather than independent held-out validation.}
\label{tab:cpg_model_fit}
\small
\setlength{\tabcolsep}{5pt}
\resizebox{\textwidth}{!}{%
\begin{tabular}{rrrrrr}
\toprule
Temperature ($^\circ$C) & DST RMSE & US06 RMSE & FUDS RMSE & Mean RMSE & Mean MAE \\
\midrule
$-10$ & 55.690 & 59.517 & 66.539 & 60.582 & 45.688 \\
$0$   & 44.250 & 41.984 & 59.736 & 48.656 & 32.098 \\
$10$  & 30.824 & 34.603 & 23.371 & 29.599 & 18.104 \\
$20$  & 36.703 & 14.559 & 25.636 & 25.633 & 13.460 \\
$25$  & 32.074 & 20.679 & 29.471 & 27.408 & 13.604 \\
$30$  & 21.564 & 19.802 & 23.376 & 21.581 & 13.738 \\
$40$  & 29.265 & 21.641 & 25.445 & 25.450 & 17.860 \\
$50$  & 26.105 & 24.618 & 26.195 & 25.639 & 19.814 \\
\midrule
Condition-macro mean & 34.559 & 29.675 & 34.971 & 33.069 & 21.796 \\
\bottomrule
\end{tabular}
}
\end{table}

\begin{revisionblock}
The supervised healthy-residual calibration was evaluated independently for every temperature--cycle target. For each entry in Table~\ref{tab:healthy_residual_calibration} at temperature $T$ and target cycle $c$, the regressor specified in \ref{app:calibration_implementation} was trained using all available healthy samples from the other two drive cycles at the same temperature. It was then evaluated on the 50\%-SOC-anchored segment of the excluded cycle. During this held-out audit only, reference SOC was supplied as the SOC feature so that the reported value measures residual-map approximation rather than SOC-estimation error. The target remained the measured healthy voltage minus the open-loop CPG-SPMT voltage.

Each table entry is the RMSE between the observed healthy-model residual and its cross-cycle prediction. Temperature means give equal weight to the three held-out cycles, cycle means give equal weight to the eight temperatures, and 9.902 mV is the equal-weight macro mean over all 24 targets. The larger errors at $-10$ and $0^\circ$C indicate stronger temperature- and excitation-dependent model discrepancy. The 9.902 mV value is not directly comparable to the 33.069 mV full-trajectory raw-model RMSE in Table~\ref{tab:cpg_model_fit} because the evaluated windows, targets, and training split differ. These are calibration-map errors, not posterior estimator voltage-fit errors or sensor-fault estimates.

\begin{table}[H]
\centering
\color{revisionyellow}
\caption{Cross-cycle held-out RMSE of the supervised healthy-model residual map on 50\%-SOC-anchored segments (mV). For each temperature and target cycle, the other two drive cycles at the same temperature form the training set. The bottom-right value is the equal-weight macro mean over the 24 held-out temperature--cycle targets.}
\label{tab:healthy_residual_calibration}
\small
\setlength{\tabcolsep}{6pt}
\begin{tabular}{rrrrr}
\toprule
Temperature ($^\circ$C) & DST & FUDS & US06 & Temperature mean \\
\midrule
$-10$ & 37.353 & 41.497 & 29.803 & 36.218 \\
$0$   & 25.291 & 25.496 & 17.996 & 22.928 \\
$10$  & 7.241  & 6.531  & 5.252  & 6.341 \\
$20$  & 4.652  & 3.376  & 2.890  & 3.639 \\
$25$  & 3.409  & 2.658  & 2.126  & 2.731 \\
$30$  & 3.144  & 2.419  & 1.710  & 2.424 \\
$40$  & 2.937  & 2.052  & 2.781  & 2.590 \\
$50$  & 2.696  & 2.336  & 1.996  & 2.343 \\
\midrule
Cycle mean & 10.841 & 10.796 & 8.069 & 9.902 \\
\bottomrule
\end{tabular}
\end{table}
\end{revisionblock}

\subsection{Overall fault-tolerant performance under bias and gain faults}
\label{subsec:overall_results}

\begin{figure}[H]
    \centering
    \includegraphics[width=\textwidth]{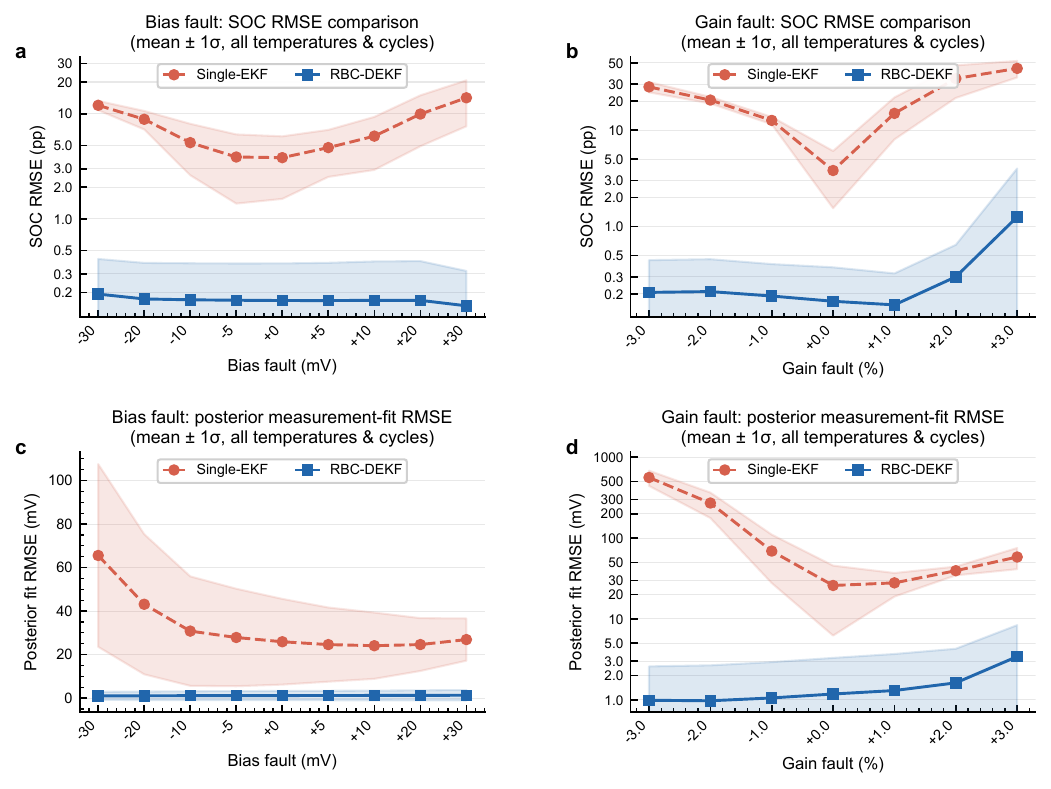}
    \caption{Overall performance comparison under additive bias faults and multiplicative gain faults. Panels (a) and (b) show the average SOC RMSE as a function of fault magnitude over all temperatures and drive cycles. Panels (c) and (d) show the corresponding \rev{posterior faulty-measurement fit RMSE}. Shaded bands denote $\pm1$ standard deviation across the matched temperature--cycle conditions, rather than repeated noise realizations. \rev{The voltage metric is not a clean-voltage recovery error.}}
    \label{fig:overview}
\end{figure}

\rev{Figure~\ref{fig:overview}, together with Tables~\ref{tab:bias_comprehensive} and~\ref{tab:gain_comprehensive}, summarizes the overall estimation performance of the \rev{original RBC-DEKF} under additive bias faults and multiplicative gain faults. The comparison uses SOC-estimation accuracy and posterior faulty-measurement fit across the full operating matrix; the latter must not be interpreted as recovery of an unobserved clean voltage.}

\begin{table}[htbp]
\centering
\color{revisionyellow}
\caption{Expanded static-fault comparison: mean SOC RMSE (pp) over the original condition--fault matrices. The oracle column subtracts the exact injected voltage residual before the Single-EKF and is therefore unavailable in practice.}
\label{tab:joint_zero_theta}
\small
\setlength{\tabcolsep}{3.5pt}
\resizebox{\textwidth}{!}{%
\begin{tabular}{lrrrrrr}
\toprule
Fault family & \makecell{Profiles per\\method} & Single-EKF & \makecell{Oracle\\removal} & \makecell{RBC-DEKF\\warm start} & \makecell{RBC-DEKF\\$\theta_0=0$} & \makecell{Conventional\\joint EKF} \\
\midrule
Additive bias & 216 & 7.646099 & 3.821 & 0.169738 & 0.169738 & \textbf{0.092910} \\
Multiplicative gain & 168 & 22.645544 & 3.821 & 0.356183 & 0.356183 & \textbf{0.092684} \\
\bottomrule
\end{tabular}
}
\end{table}

\begin{revisionblock}
Table~\ref{tab:joint_zero_theta} compares residual-state augmentation under the accurate initial states used in the static matrix. The conventional joint EKF has the lowest mean SOC RMSE and is better than the \rev{original RBC-DEKF} in 276 of 384 paired condition--fault profiles. The oracle direct-removal reference restores the no-fault Single-EKF result (3.821 percentage points), whereas both residual-state estimators also accommodate systematic model discrepancy.

Accordingly, the oracle control answers the narrower question of what happens when the injected sensor abnormality is removed exactly; the residual-state estimators also compensate systematic model discrepancy already present in the measured data.

Initializing the residual state at zero is practically equivalent to the former warm start. The mean absolute SOC-RMSE difference is below $5\times10^{-7}$ percentage points in each fault family. With the inherited settings,
\begin{equation}
K_{\theta,0}=\frac{P_{\theta,0}+Q_\theta}{P_{\theta,0}+Q_\theta+R_\theta}
\approx0.9999999,
\label{eq:first_residual_gain}
\end{equation}
so the first online residual update rapidly suppresses the initialization difference. Because the state update precedes the residual update, the two initializations need not produce identical first-step state estimates. All subsequent main results therefore use $\hat\theta_0=0$ and do not rely on a preconfigured first-sample residual.
\end{revisionblock}

\rev{For additive bias faults, the \rev{original RBC-DEKF} exhibits stable behavior over the full fault range. Across the 216 profiles in Table~\ref{tab:bias_comprehensive}, SOC RMSE decreases from 7.646 pp for the Single-EKF to 0.170 pp for the RBC-DEKF, while posterior measurement-fit RMSE decreases from 32.617 to 1.179 mV. At individual bias levels, the baseline SOC RMSE increases from 3.821 pp at 0 mV to 14.156 pp at $+30$ mV, whereas the RBC-DEKF remains between 0.149 and 0.193 pp. Figure~\ref{fig:overview}(a) and (c) show the same uniform response, consistent with the residual loop accommodating a predominantly additive voltage discrepancy.}

\rev{For the 168 multiplicative-gain profiles, average SOC RMSE decreases from 22.646 to 0.356 pp and average posterior measurement-fit RMSE decreases from 150.382 to 1.505 mV. Thus, the scalar residual state accommodates the voltage-domain effect of the gain abnormality without explicitly estimating its coefficient.}

\rev{At similar equivalent voltage magnitudes, the $-1\%$ and $+1\%$ gain faults produce average residuals of $-31.660$ and $+31.660$ mV. Their compensated SOC RMSE values are 0.190 and 0.154 pp, close to the 0.193 and 0.149 pp values under $-30$ and $+30$ mV additive biases. The corresponding fit RMSE pairs are likewise close (1.054 versus 1.037 mV and 1.302 versus 1.320 mV).}

\rev{The $\pm2\%$ and $\pm3\%$ gain faults create average equivalent residual magnitudes of approximately 63.321 and 94.981 mV, respectively, and also vary with terminal voltage. These severe cases therefore demand larger and faster residual correction than the tested $\pm30$ mV offsets. At $+3\%$, the 24-condition averages are 1.261 pp in SOC RMSE and 3.425 mV in fit RMSE, compared with 43.872 pp and 58.305 mV for the Single-EKF.}

\rev{Gain-fault direction affects the two baseline metrics differently: the Single-EKF SOC RMSE reaches 43.872 pp at $+3\%$, whereas its fit RMSE reaches 271.040 and 561.030 mV at $-2\%$ and $-3\%$. The RBC-DEKF suppresses both responses.}

Overall, Figure~\ref{fig:overview} and Tables~\ref{tab:bias_comprehensive}--\ref{tab:gain_comprehensive} show that the \rev{original RBC-DEKF} provides effective compensation for both additive and multiplicative voltage abnormalities. Additive offsets yield the most uniform performance because their equivalent residual is approximately constant. Gain faults are also accurately compensated over most of the tested domain. The remaining degradation is concentrated at the largest equivalent residuals and most demanding voltage trajectories.

\begin{table}[H]
\centering
\caption{Comprehensive comparison between the Single-EKF and the \rev{original RBC-DEKF} under additive bias faults. Values are means over matched conditions. SOC RMSE is reported in pp, and \rev{posterior faulty-measurement fit RMSE} is reported in mV.}
\label{tab:bias_comprehensive}
\renewcommand{\arraystretch}{1.05}
\setlength{\tabcolsep}{2pt}
\scriptsize
\resizebox{\textwidth}{!}{%
\begin{tabular}{lccccc}
\toprule
\textbf{Group / Condition} & \shortstack{\textbf{Single-EKF}\\\textbf{SOC RMSE (pp)}} & \shortstack{\rev{\textbf{RBC-DEKF}}\\\textbf{SOC RMSE (pp)}} & \shortstack{\textbf{Single-EKF}\\\textbf{Fit RMSE (mV)}} & \shortstack{\rev{\textbf{RBC-DEKF}}\\\textbf{Fit RMSE (mV)}} & \textbf{Cases} \\
\midrule
\multicolumn{6}{l}{\textbf{Overall}} \\
Overall average & 7.646 & 0.170 & 32.617 & 1.179 & 216 \\
\midrule
\multicolumn{6}{l}{\textbf{By Bias Fault Level (mV)}} \\
$-30$ & 12.010 & 0.193 & 65.570 & 1.037 & 24 \\
$-20$ & 8.848  & 0.174 & 43.144 & 1.078 & 24 \\
$-10$ & 5.307  & 0.170 & 30.807 & 1.127 & 24 \\
$-5$  & 3.875  & 0.169 & 27.876 & 1.153 & 24 \\
$0$   & 3.821  & 0.168 & 25.916 & 1.179 & 24 \\
$+5$  & 4.761  & 0.168 & 24.608 & 1.206 & 24 \\
$+10$ & 6.120  & 0.168 & 24.096 & 1.233 & 24 \\
$+20$ & 9.917  & 0.168 & 24.622 & 1.275 & 24 \\
$+30$ & 14.156 & 0.149 & 26.910 & 1.320 & 24 \\
\midrule
\multicolumn{6}{l}{\textbf{By Temperature ($^\circ$C)}} \\
$-10$ & 7.715 & 0.169 & 31.631 & 5.549 & 27 \\
$0$   & 6.414 & 0.162 & 25.667 & 2.616 & 27 \\
$10$  & 6.515 & 0.157 & 25.980 & 0.371 & 27 \\
$20$  & 7.041 & 0.163 & 28.562 & 0.220 & 27 \\
$25$  & 7.431 & 0.157 & 31.172 & 0.128 & 27 \\
$30$  & 7.984 & 0.160 & 34.731 & 0.125 & 27 \\
$40$  & 8.835 & 0.175 & 39.666 & 0.153 & 27 \\
$50$  & 9.233 & 0.213 & 43.525 & 0.267 & 27 \\
\midrule
\multicolumn{6}{l}{\textbf{By Drive Cycle}} \\
DST  & 7.654 & 0.162 & 32.492 & 1.149 & 72 \\
FUDS & 7.384 & 0.145 & 32.529 & 1.419 & 72 \\
US06 & 7.900 & 0.202 & 32.828 & 0.968 & 72 \\
\bottomrule
\end{tabular}
}
\end{table}

\begin{table}[H]
\centering
\caption{Comprehensive comparison between the Single-EKF and the \rev{original RBC-DEKF} under multiplicative gain faults. Values are means over matched conditions. SOC RMSE is reported in pp, and \rev{posterior faulty-measurement fit RMSE} is reported in mV. The average gain-induced equivalent voltage residual is shown in parentheses for each gain level.}
\label{tab:gain_comprehensive}
\renewcommand{\arraystretch}{1.05}
\setlength{\tabcolsep}{2pt}
\scriptsize
\resizebox{\textwidth}{!}{%
\begin{tabular}{lccccc}
\toprule
\textbf{Group / Condition} & \shortstack{\textbf{Single-EKF}\\\textbf{SOC RMSE (pp)}} & \shortstack{\rev{\textbf{RBC-DEKF}}\\\textbf{SOC RMSE (pp)}} & \shortstack{\textbf{Single-EKF}\\\textbf{Fit RMSE (mV)}} & \shortstack{\rev{\textbf{RBC-DEKF}}\\\textbf{Fit RMSE (mV)}} & \textbf{Cases} \\
\midrule
\multicolumn{6}{l}{\textbf{Overall}} \\
Overall average & 22.646 & 0.356 & 150.382 & 1.505 & 168 \\
\midrule
\multicolumn{6}{l}{\textbf{By Gain Fault Level (\% / average equivalent voltage residual)}} \\
$-3\%$ ($-94.981$ mV) & 28.206 & 0.207 & 561.030 & 0.986 & 24 \\
$-2\%$ ($-63.321$ mV) & 20.572 & 0.212 & 271.040 & 0.972 & 24 \\
$-1\%$ ($-31.660$ mV) & 12.656 & 0.190 & 69.001  & 1.054 & 24 \\
$0\%$ ($0$ mV)        & 3.821  & 0.168 & 25.916  & 1.179 & 24 \\
$+1\%$ ($+31.660$ mV) & 14.975 & 0.154 & 27.923  & 1.302 & 24 \\
$+2\%$ ($+63.321$ mV) & 34.417 & 0.301 & 39.462  & 1.620 & 24 \\
$+3\%$ ($+94.981$ mV) & 43.872 & 1.261 & 58.305  & 3.425 & 24 \\
\midrule
\multicolumn{6}{l}{\textbf{By Temperature ($^\circ$C)}} \\
$-10$ & 13.725 & 0.669 & 111.838 & 5.704 & 21 \\
$0$   & 18.989 & 0.463 & 180.467 & 2.831 & 21 \\
$10$  & 22.534 & 0.203 & 184.055 & 0.408 & 21 \\
$20$  & 24.608 & 0.774 & 161.771 & 1.093 & 21 \\
$25$  & 24.780 & 0.328 & 154.576 & 0.804 & 21 \\
$30$  & 25.084 & 0.270 & 142.669 & 0.623 & 21 \\
$40$  & 25.595 & 0.090 & 133.170 & 0.296 & 21 \\
$50$  & 25.849 & 0.053 & 134.512 & 0.284 & 21 \\
\midrule
\multicolumn{6}{l}{\textbf{By Drive Cycle}} \\
DST  & 22.314 & 0.187 & 151.277 & 1.282 & 56 \\
FUDS & 22.711 & 0.240 & 151.643 & 1.469 & 56 \\
US06 & 22.912 & 0.641 & 148.226 & 1.766 & 56 \\
\bottomrule
\end{tabular}
}
\end{table}

\subsection{Representative results under additive bias faults}
\label{subsec:bias_results}

\begin{figure}[H]
    \centering
    \includegraphics[width=\textwidth]{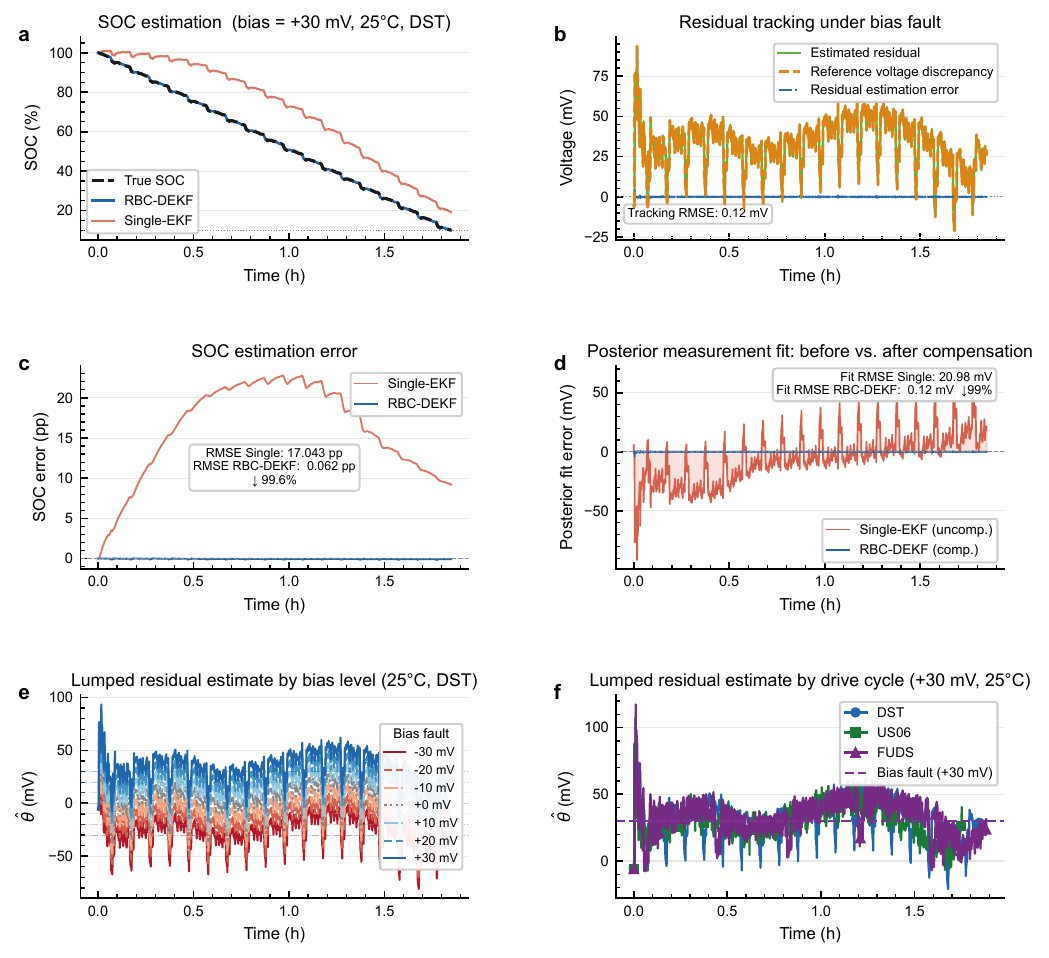}
    \caption{Representative results under additive bias fault. Panels (a)--(d) show the SOC trajectories, residual tracking, SOC error, and \rev{posterior faulty-measurement fit error} before and after compensation for the case of $+30$ mV at $25^\circ\mathrm{C}$ under DST. Panels (e) and (f) show the evolution of the estimated lumped voltage discrepancy across different bias levels and across different drive cycles.}
    \label{fig:bias_case}
\end{figure}

\rev{Figure~\ref{fig:bias_case} presents a representative additive-bias case at $25^\circ\mathrm{C}$ under the DST cycle with an injected bias of $+30$ mV, while Table~\ref{tab:representative_cases} summarizes the corresponding severe-fault results at $25^\circ\mathrm{C}$ for all three drive cycles. For this representative case, the SOC RMSE decreases from 17.043 to 0.062 pp, and the posterior faulty-measurement fit RMSE decreases from 20.98 to 0.12 mV. The Single-EKF deviates markedly from the reference SOC trajectory, whereas the RBC-DEKF remains close to it.}

\rev{In Figures~\ref{fig:bias_case}--\ref{fig:gain_case}, the reference voltage discrepancy is $z_k-h(\hat{\mathbf{x}}_k,I_k,T_k)$: the faulty measurement minus the uncompensated model voltage after the state update. It is a posterior computational reference, not sensor-fault ground truth. Figure~\ref{fig:bias_case}(b) shows the voltage-discrepancy state tracking this combined offset, model discrepancy, and measurement noise, consistent with the small posterior fit error in panel (d).}

\rev{At $25^\circ\mathrm{C}$ and $+30$ mV bias, the Single-EKF SOC RMSE is 17.043, 16.307, and 17.315 pp under DST, US06, and FUDS, respectively, whereas the \rev{original RBC-DEKF} reduces these values to 0.062, 0.060, and 0.075 pp. Figure~\ref{fig:bias_case}(e)--(f) show that the estimated lumped residual follows the injected offset while retaining the condition-dependent model discrepancy.}

\Needspace{5\baselineskip}
\subsection{Performance under multiplicative gain faults}
\label{subsec:gain_results}

\begin{figure}[H]
    \centering
    \includegraphics[width=\textwidth]{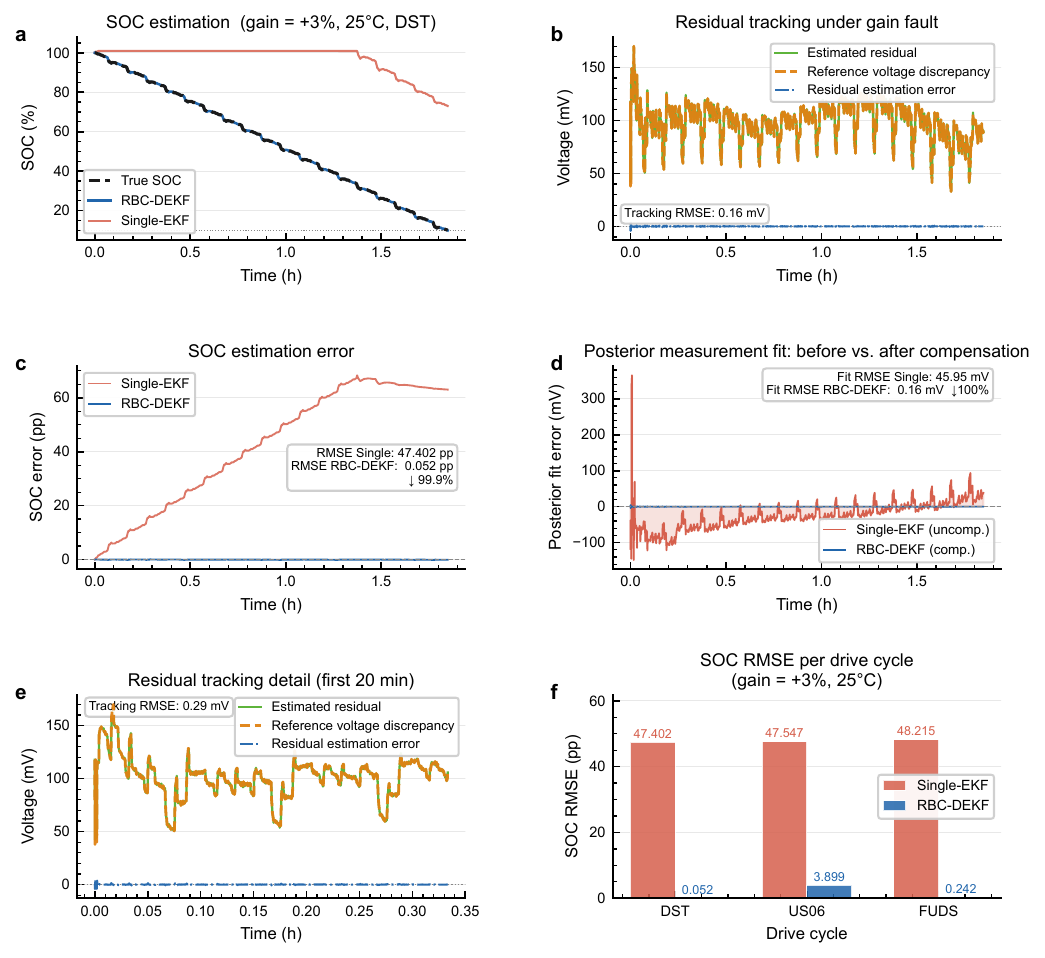}
    \caption{Representative results under multiplicative gain fault. Panels (a)--(d) show the SOC trajectories, residual tracking, SOC error, and \rev{posterior faulty-measurement fit error} before and after compensation for the case of $+3\%$ at $25^\circ\mathrm{C}$ under DST. Panel (e) shows the residual tracking detail during the first 20 minutes, and panel (f) compares the SOC RMSE across drive cycles.}
    \label{fig:gain_case}
\end{figure}

\rev{Figure~\ref{fig:gain_case} shows a representative $+3\%$ gain-fault case at $25^\circ\mathrm{C}$ under DST. The SOC RMSE decreases from 47.402 to 0.052 pp, and the posterior faulty-measurement fit RMSE decreases from 45.95 to 0.16 mV. The Single-EKF develops a rapidly increasing SOC deviation, whereas the RBC-DEKF remains close to the reference trajectory.}

\rev{Although the underlying abnormality is multiplicative, its effect on the voltage channel contributes the scalar equivalent residual $\alpha V_k^{\mathrm{true}}$. Figure~\ref{fig:gain_case}(b) and (e) show close tracking of the reference voltage discrepancy, which contains the gain-induced contribution, model discrepancy, and measurement noise. For the representative DST case, the average equivalent gain-induced contribution is 95.838 mV, yet the posterior measurement-fit RMSE is 0.16 mV. Explicit identification of the gain coefficient is not required for the reported voltage-domain accommodation in this case.}

\rev{The compensated SOC RMSE is 0.052 and 0.242 pp under DST and FUDS, respectively. Under US06 it increases to 3.899 pp, still a 91.800\% reduction relative to the Single-EKF value of 47.547 pp. The corresponding posterior measurement-fit RMSE is 10.20 mV, reflecting the larger tracking demand of an equivalent residual near 95.6 mV.}

\rev{The final column of Table~\ref{tab:representative_cases} reports the posterior faulty-measurement fit RMSE for all three drive cycles; the corresponding Single-EKF fit errors for the representative DST cases are shown explicitly in Figures~\ref{fig:bias_case}(d) and~\ref{fig:gain_case}(d).}

\begin{table}[H]
\centering
\caption{Representative SOC-estimation results and \rev{posterior faulty-measurement fit RMSE} at $25^\circ\mathrm{C}$ under severe additive and multiplicative faults.}
\label{tab:representative_cases}
\renewcommand{\arraystretch}{1.1}
\setlength{\tabcolsep}{2pt}
\scriptsize
\begin{tabular*}{\textwidth}{@{\extracolsep{\fill}}cccccccc}
\toprule
\textbf{Type} & \textbf{Level} & \textbf{Cycle} & \shortstack{\textbf{Mean equiv.}\\\textbf{residual (mV)}} & \shortstack{\textbf{Single SOC}\\\textbf{RMSE (pp)}} & \shortstack{\rev{\textbf{RBC-DEKF SOC}}\\\textbf{RMSE (pp)}} & \shortstack{\textbf{Improve.}\\\textbf{(\%)}} & \shortstack{\rev{\textbf{Posterior fit}}\\\textbf{RMSE (mV)}} \\
\midrule
Bias & $+30$ mV & DST  & --      & 17.043 & 0.062 & 99.636 & 0.12 \\
Bias & $+30$ mV & US06 & --      & 16.307 & 0.060 & 99.634 & 0.10 \\
Bias & $+30$ mV & FUDS & --      & 17.315 & 0.075 & 99.565 & 0.10 \\
\midrule
Gain & $+3\%$ & DST  & 95.838 & 47.402 & 0.052 & 99.891 & 0.16 \\
Gain & $+3\%$ & US06 & 95.628 & 47.547 & 3.899 & 91.800 & 10.20 \\
Gain & $+3\%$ & FUDS & 95.809 & 48.215 & 0.242 & 99.498 & 0.20 \\
\bottomrule
\end{tabular*}
\end{table}

\subsection{Robustness across temperatures and dynamic drive cycles}
\label{subsec:robustness_results}

\begin{figure}[H]
    \centering
    \includegraphics[width=\textwidth]{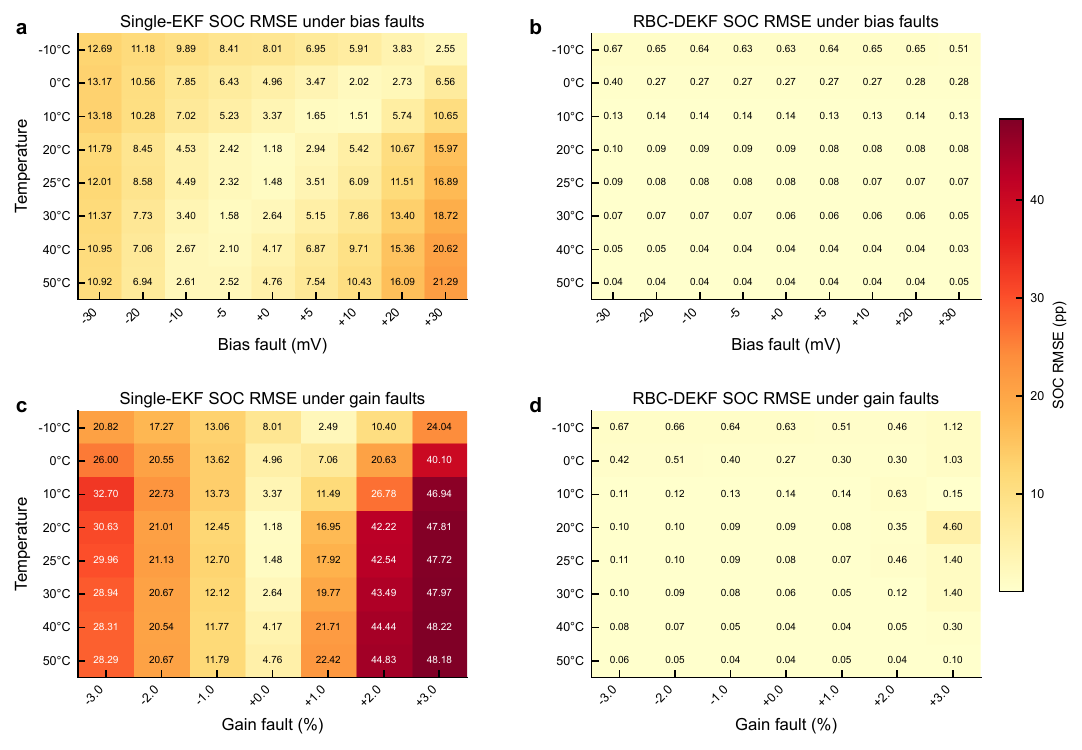}
    \caption{Heatmap representation of SOC-estimation robustness across temperatures and fault magnitudes. Each heatmap entry is the mean SOC RMSE over the three drive cycles. Panels (a) and (b) show the SOC RMSE distributions of the Single-EKF and the \rev{original RBC-DEKF} under additive bias faults, respectively. Panels (c) and (d) show the corresponding SOC RMSE distributions under multiplicative gain faults.}
    \label{fig:temp_robustness}
\end{figure}

\begin{figure}[H]
    \centering
    \includegraphics[width=\textwidth]{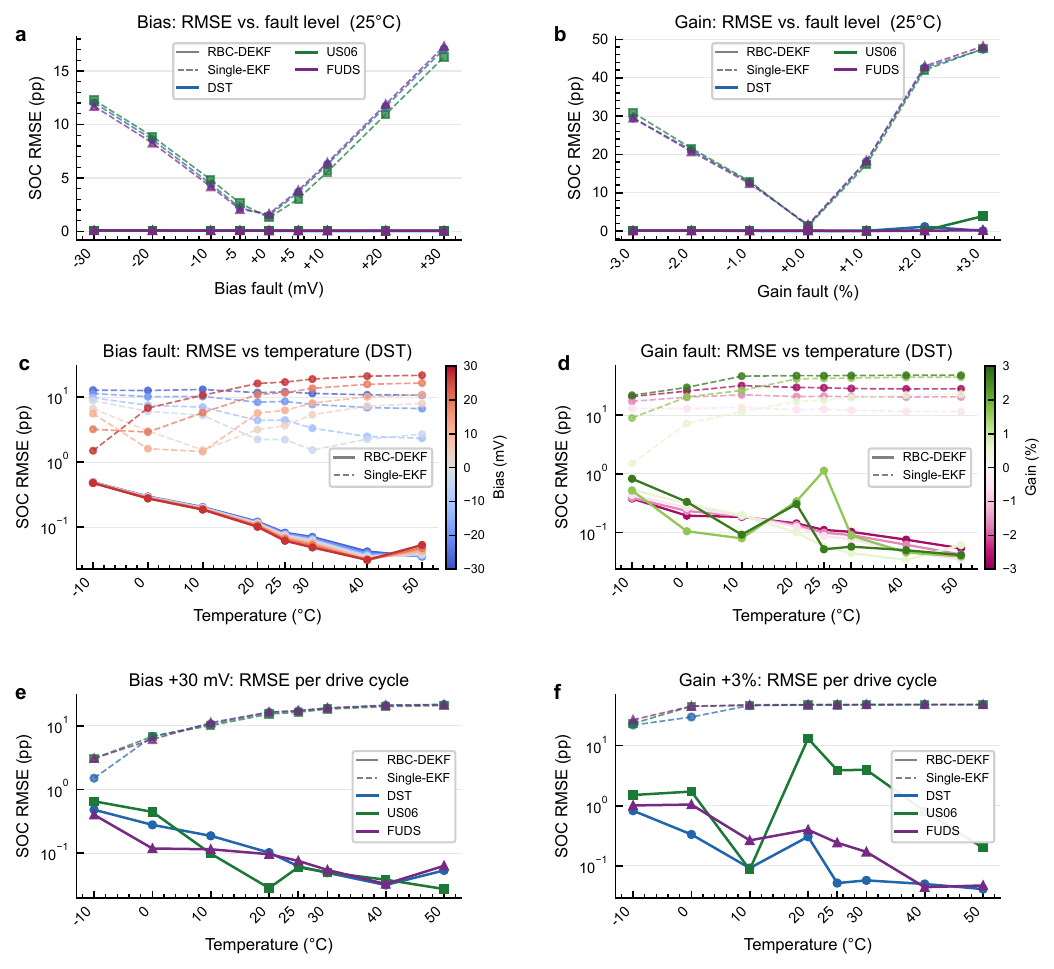}
    \caption{Robustness analysis of the \rev{original RBC-DEKF} across fault levels, temperatures, and drive cycles. Panels (a) and (b) show the SOC RMSE as a function of fault magnitude at $25^\circ\mathrm{C}$ for the three drive cycles. Panels (c) and (d) show the temperature dependence across all tested additive-bias and multiplicative-gain levels, respectively, under the DST cycle; the color bars encode the corresponding fault levels. Panels (e) and (f) compare the temperature-dependent SOC RMSE across drive cycles under the severe $+30$ mV additive-bias fault and the severe $+3\%$ multiplicative-gain fault, respectively.}
    \label{fig:robustness}
\end{figure}

The robustness of the \rev{original RBC-DEKF} was further evaluated across wide temperature variation and different dynamic drive cycles under additive and multiplicative voltage abnormalities. To provide both condition-map and aggregated-trend perspectives, the analysis is presented in two complementary forms. Figure~\ref{fig:temp_robustness} uses SOC-RMSE heatmaps to show how the error distribution changes jointly with temperature and fault magnitude for both the Single-EKF and the \rev{original RBC-DEKF}. Figure~\ref{fig:robustness} then summarizes the corresponding trends across fault levels, temperatures, and drive cycles using grouped curves. The associated numerical results are reported in Tables~\ref{tab:temp_severe_summary} and~\ref{tab:cycle_severe_summary}.

\rev{For additive bias faults, Figure~\ref{fig:temp_robustness}(a)--(b) show that the RBC-DEKF keeps SOC RMSE below 0.7 pp throughout the tested temperature--fault plane; most points are at or below 0.1 pp above $10^\circ\mathrm{C}$, whereas the Single-EKF error increases sharply with bias magnitude.}

The aggregated results in Figure~\ref{fig:robustness}(a), (c), and (e) and Table~\ref{tab:temp_severe_summary} show the same trend. Under $+30$ mV bias, compensated SOC RMSE remains below 0.51 pp at every temperature and below 0.1 pp from $20^\circ\mathrm{C}$ upward. The cycle-wise means are 0.155, 0.119, and 0.174 pp under DST, FUDS, and US06, respectively.

Under gain faults, Figure~\ref{fig:temp_robustness}(c)--(d) show low RBC-DEKF error over most of the tested plane. At $+3\%$ gain, SOC RMSE reaches 4.604 pp at $20^\circ\mathrm{C}$ but remains at or below 1.398 pp at the other temperatures. This case corresponds to an average equivalent residual near 95 mV, substantially beyond the 30 mV additive-bias range.

The aggregated fault-level and cycle-dependent trends in Figure~\ref{fig:robustness}(b), (d), and (f) further clarify this result. At $25^\circ\mathrm{C}$, the \rev{original RBC-DEKF} remains accurate over moderate gain levels and shows increased error mainly under the most severe positive gain condition. The temperature-dependent comparison in Figure~\ref{fig:robustness}(d) confirms that the remaining error is concentrated in a limited set of operating conditions rather than distributed across the full thermal range. In the voltage-residual domain, the fixed random-walk loop must track an equivalent residual whose amplitude and temporal variation both increase with the gain level and the terminal-voltage trajectory.

For $+3\%$ gain, the compensated SOC RMSE is 0.220, 0.403, and 3.161 pp under DST, FUDS, and US06, respectively. Under US06, this remains a 92.934\% mean case-wise improvement over the eight temperatures relative to the Single-EKF result of 44.227 pp.

\rev{Overall, Figures~\ref{fig:temp_robustness} and~\ref{fig:robustness}, together with Tables~\ref{tab:temp_severe_summary} and~\ref{tab:cycle_severe_summary}, show broad in-matrix robustness across temperature and drive-cycle variations for both fault families. Additive-bias performance remains nearly invariant over the test matrix. Gain-fault performance is also close to the nominal estimation level in most conditions, with the remaining errors concentrated at the largest equivalent-residual magnitude and the most demanding operating profiles. These results characterize the tested tracking range of the fixed voltage-discrepancy-loop tuning; they do not establish out-of-sample generalization.}

\begin{table}[H]
\centering
\caption{SOC estimation performance under severe additive and multiplicative faults across different temperatures (mean over three drive cycles). The improvement column is the mean of the case-wise relative RMSE reductions over the three drive cycles and therefore is not generally equal to the reduction calculated from the two aggregated RMSE values.}
\label{tab:temp_severe_summary}
\renewcommand{\arraystretch}{1.05}
\setlength{\tabcolsep}{2pt}
\scriptsize
\resizebox{\textwidth}{!}{%
\begin{tabular}{ccccccc}
\toprule
\textbf{Temperature} & \multicolumn{3}{c}{$+30$ mV bias fault} & \multicolumn{3}{c}{$+3\%$ gain fault} \\
\cmidrule(lr){2-4} \cmidrule(lr){5-7}
\textbf{($^\circ$C)} & \shortstack{\textbf{Single}\\\textbf{RMSE (pp)}} & \shortstack{\rev{\textbf{RBC-DEKF}}\\\textbf{RMSE (pp)}} & \shortstack{\textbf{Mean improv.}\\\textbf{(\%)}} & \shortstack{\textbf{Single}\\\textbf{RMSE (pp)}} & \shortstack{\rev{\textbf{RBC-DEKF}}\\\textbf{RMSE (pp)}} & \shortstack{\textbf{Mean improv.}\\\textbf{(\%)}} \\
\midrule
$-10$ & 2.547  & 0.509 & 77.974 & 24.041 & 1.115 & 95.334 \\
$0$   & 6.557  & 0.279 & 95.820 & 40.098 & 1.033 & 97.587 \\
$10$  & 10.653 & 0.133 & 98.754 & 46.937 & 0.148 & 99.687 \\
$20$  & 15.967 & 0.075 & 99.535 & 47.811 & 4.604 & 90.352 \\
$25$  & 16.888 & 0.066 & 99.612 & 47.721 & 1.398 & 97.063 \\
$30$  & 18.723 & 0.051 & 99.727 & 47.968 & 1.396 & 97.081 \\
$40$  & 20.622 & 0.034 & 99.834 & 48.218 & 0.301 & 99.373 \\
$50$  & 21.287 & 0.048 & 99.776 & 48.182 & 0.096 & 99.800 \\
\bottomrule
\end{tabular}
}
\end{table}

\begin{table}[H]
\centering
\caption{SOC estimation performance under severe additive and multiplicative faults across different drive cycles (mean over all temperatures). The improvement column is the mean of the case-wise relative RMSE reductions over the eight temperatures and therefore is not generally equal to the reduction calculated from the two aggregated RMSE values.}
\label{tab:cycle_severe_summary}
\renewcommand{\arraystretch}{1.05}
\setlength{\tabcolsep}{2pt}
\scriptsize
\resizebox{\textwidth}{!}{%
\begin{tabular}{ccccccc}
\toprule
\textbf{Cycle} & \multicolumn{3}{c}{$+30$ mV bias fault} & \multicolumn{3}{c}{$+3\%$ gain fault} \\
\cmidrule(lr){2-4} \cmidrule(lr){5-7}
 & \shortstack{\textbf{Single}\\\textbf{RMSE (pp)}} & \shortstack{\rev{\textbf{RBC-DEKF}}\\\textbf{RMSE (pp)}} & \shortstack{\textbf{Mean improv.}\\\textbf{(\%)}} & \shortstack{\textbf{Single}\\\textbf{RMSE (pp)}} & \shortstack{\rev{\textbf{RBC-DEKF}}\\\textbf{RMSE (pp)}} & \shortstack{\textbf{Mean improv.}\\\textbf{(\%)}} \\
\midrule
DST  & 14.233 & 0.155 & 95.080 & 42.166 & 0.220 & 99.232 \\
FUDS & 14.422 & 0.119 & 97.790 & 45.222 & 0.403 & 98.938 \\
US06 & 13.812 & 0.174 & 96.267 & 44.227 & 3.161 & 92.934 \\
\bottomrule
\end{tabular}
}
\end{table}

\subsection{State recovery and fault tolerance under initial-SOC uncertainty}
\label{subsec:initial_soc_recovery}

\begin{revisionblock}
This subsection evaluates recovery from initialization uncertainty; accurately initialized static-fault performance has already been evaluated in Sections~\ref{subsec:overall_results}--\ref{subsec:robustness_results}. Throughout this subsection and Section~\ref{subsec:delayed_faults}, \emph{commissioned RBC-DEKF} denotes the complete realization of Section~\ref{subsec:commissioning}, including the frozen cross-cycle calibration retained after handover. The paired stress test uses 50\%-SOC-anchored segments from three measured trajectories ($0^\circ$C--US06, $25^\circ$C--DST, and $50^\circ$C--FUDS) and imposes initial errors of $-20$, $-10$, $-5$, $+5$, $+10$, and $+20$ percentage points. Each reported value is a macro mean over these three segments. The six offsets reuse each segment and noise realization and are sensitivity conditions rather than independent experiments.
\end{revisionblock}

\begin{table}[htbp]
\centering
\color{revisionyellow}
\caption{Paired initial-SOC sensitivity: full-record macro-mean SOC RMSE (pp) over three representative 50\%-SOC-anchored segments. The same complete recorded interval, including startup provisional outputs, is used for all four methods.}
\label{tab:initial_soc_recovery}
\small
\setlength{\tabcolsep}{4pt}
\resizebox{\textwidth}{!}{%
\begin{tabular}{lrrrr}
\toprule
\makecell{Initial error\\(pp)} & \makecell{Open-loop\\CPG-SPMT} & \makecell{Original\\RBC-DEKF} & \makecell{Conventional\\joint EKF} & \makecell{Commissioned\\RBC-DEKF} \\
\midrule
$-20$ & 19.069 & 18.058 & 18.816 & \textbf{1.683} \\
$-10$ & 10.052 &  9.421 &  9.791 & \textbf{1.286} \\
$-5$  &  5.052 &  4.652 &  4.909 & \textbf{1.238} \\
$+5$  &  4.948 &  5.056 &  4.978 & \textbf{1.371} \\
$+10$ &  9.948 &  9.986 &  9.957 & \textbf{2.219} \\
$+20$ & 19.948 & 19.976 & 19.950 & \textbf{2.681} \\
\midrule
Mean & 11.503 & 11.192 & 11.400 & \textbf{1.747} \\
\bottomrule
\end{tabular}
}
\end{table}

\begin{figure}[htbp]
    \centering
    \includegraphics[width=\textwidth]{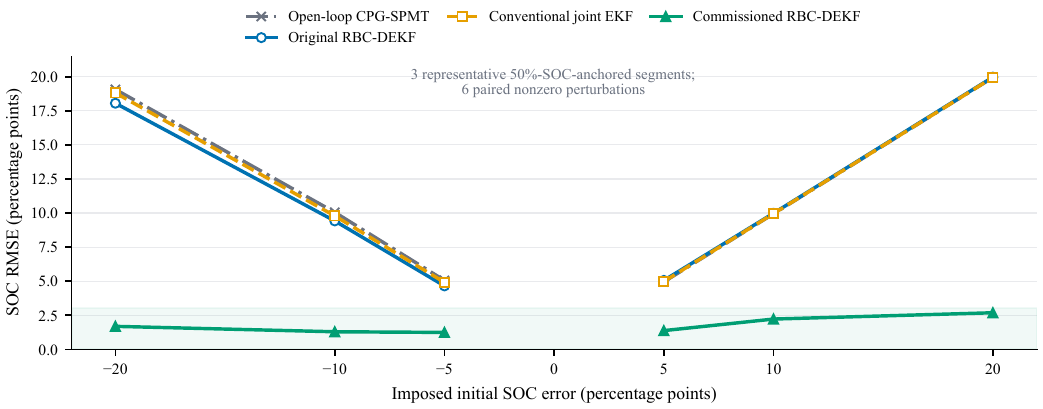}
    \caption{\rev{Full-record initial-SOC sensitivity for six nonzero perturbations. Each point averages the complete recorded SOC-error sequence, including startup, over three representative 50\%-SOC-anchored segments. The zero-error control is reported separately and is excluded from the plot.}}
    \label{fig:initial_soc_recovery}
\end{figure}

\begin{figure}[H]
    \centering
    \includegraphics[width=\textwidth]{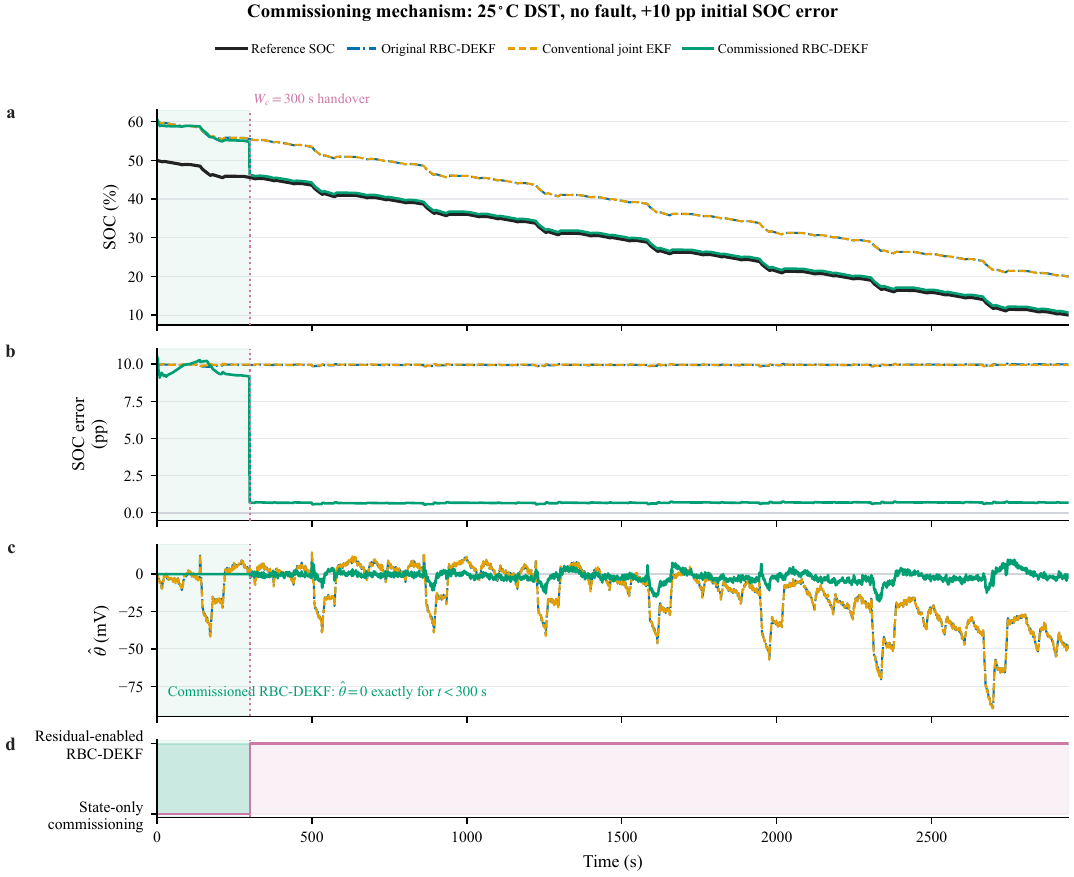}
    \caption{\rev{Causal commissioning behavior for the $25^\circ$C--DST no-fault trajectory with a $+10$-percentage-point initial-SOC perturbation. Panels (a)--(c) compare the original RBC-DEKF, conventional always-on joint EKF, and commissioned RBC-DEKF in SOC, SOC error, and lumped residual state; panel (d) shows the commissioned estimator mode. Its residual state is suspended at zero for $t<300$ s, and window replay does not backfill previously emitted outputs.}}
    \label{fig:commissioning_mechanism}
\end{figure}

\begin{revisionblock}
Table~\ref{tab:initial_soc_recovery} and Figure~\ref{fig:initial_soc_recovery} include the startup transient in every method's RMSE. The complete commissioned realization reduces the mean from 11.192 pp for the original RBC-DEKF and 11.400 pp for the joint EKF to 1.747 pp. This evaluates the complete operational realizations rather than isolating the sequential-versus-joint architecture. The separate zero-error controls give 0.057, 0.349, 0.144, and 1.235 pp for open-loop CPG-SPMT, original RBC-DEKF, conventional joint EKF, and commissioned RBC-DEKF, respectively. Thus commissioning carries a nominal-accuracy cost and is reserved for an uncertain initial prior. The calibration ablation in \ref{app:commissioning_ablation} uses the same full-record metric and gives 8.665 pp without calibration versus 1.747 pp with calibration.

The mechanism is illustrated in Figure~\ref{fig:commissioning_mechanism} for the $25^\circ$C--DST trajectory with a $+10$-percentage-point initial error and no injected voltage fault. The original always-on RBC-DEKF and conventional joint EKF assign much of the startup discrepancy to $\hat\theta$, leaving post-300-s SOC RMSE values of 9.940 and 9.946 percentage points, respectively. In the complete calibrated commissioned realization, $\hat\theta$ remains zero during the verified-healthy window; calibrated least-squares correction, causal state-EKF replay, and handover reduce the error to 0.671 percentage points. The pre-handover curve is the causal provisional output and is not retroactively replaced.
\end{revisionblock}

\subsubsection{Commissioning-window sensitivity}
\label{subsec:window_sensitivity}

\rev{The correction error is defined as the selected SOC-aligned correction minus the correction required to cancel the imposed initial-SOC perturbation, with both expressed in percentage points. Thus, an imposed $+10$-pp initial error requires a $-10$-pp correction. The correction RMSE is the root-mean-square of these errors over the 18 paired nonzero-error profiles.}

\begin{table}[htbp]
\centering
\color{revisionyellow}
\caption{Commissioning-window sensitivity for the retained bounded-$L_2$ search over three representative trajectories. Correction, common-horizon, and tail RMSE are calculated over 18 paired nonzero-initial-error runs; the nonzero-correction rate for zero-error initialization is calculated over three zero-error runs. Common-horizon RMSE uses $t\geq600\,\mathrm{s}$ for every window.}
\label{tab:window_sensitivity}
\scriptsize
\setlength{\tabcolsep}{3pt}
\resizebox{\textwidth}{!}{%
\begin{tabular}{rrrrr}
\toprule
$W_c$ (s) &
\makecell{Correction\\RMSE (pp)} &
\makecell{Common-horizon\\RMSE (pp)} &
\makecell{Tail\\RMSE (pp)} &
\makecell{Nonzero correction in\\zero-error cases (\%)} \\
\midrule
60  & 2.192 & 2.064 & 2.102 & 66.7 \\
120 & 1.431 & 1.864 & 1.900 & 33.3 \\
180 & \textbf{0.277} & 1.044 & 1.053 & 33.3 \\
300 & \textbf{0.277} & 1.122 & 1.141 & 33.3 \\
600 & \textbf{0.277} & \textbf{0.965} & \textbf{0.974} & 33.3 \\
\bottomrule
\end{tabular}
}
\end{table}

\begin{figure}[htbp]
    \centering
    \includegraphics[width=\textwidth]{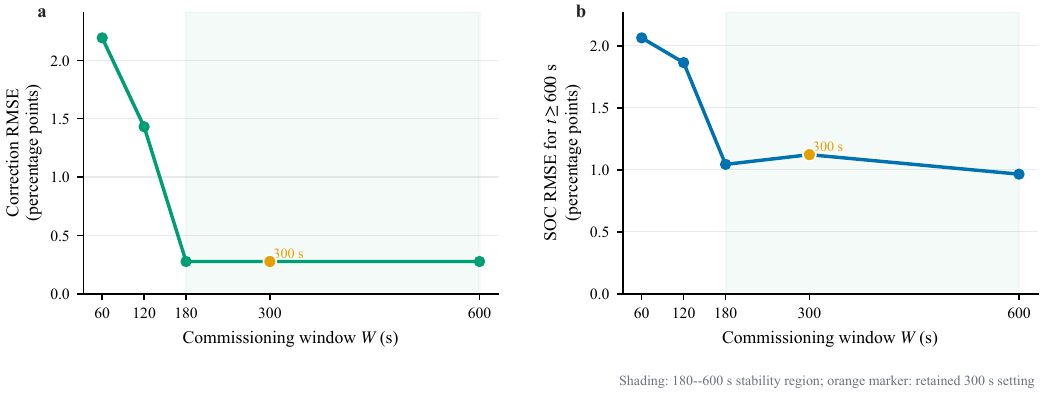}
    \caption{\rev{Commissioning-window sensitivity. (a) SOC-aligned correction RMSE; (b) commissioned RBC-DEKF SOC RMSE over the common $t\geq600\,\mathrm{s}$ evaluation horizon. Shading denotes the 180--600 s range over which the correction RMSE remains at its observed minimum level, and the orange point marks the retained 300 s setting.}}
    \label{fig:window_sensitivity}
\end{figure}

\begin{revisionblock}
The correction-error RMSE decreases from 2.192 percentage points at 60 s and 1.431 percentage points at 120 s to 0.277 percentage points at 180 s, and remains at 0.277 percentage points for the 300- and 600-s windows. Over the common $t\geq600$ s evaluation horizon, the commissioned-estimator RMSE ranges from 0.965 to 1.122 percentage points for the 180--600 s windows. Thus, on the tested trajectories, the SOC-aligned correction accuracy reaches a stable level for windows of approximately 180 s or longer, while the subsequent estimator accuracy remains similar across these longer-window settings.

The sensitivity study is used to assess the robustness of the retained commissioning window rather than to optimize it. Windows of 60 and 120 s do not yet reach the stable correction accuracy observed for the longer windows, whereas 180, 300, and 600 s yield the same correction RMSE of 0.277 pp. The 180-s case is the earliest tested window at which this stable correction level is reached. Extending the window from 300 to 600 s does not further improve the correction RMSE, while doubling the commissioning duration. We therefore retain 300 s as a practical interior setting within the observed 180--600-s stable-correction range, avoiding selection directly at the lower boundary while not incurring the full delay of the 600-s case. This choice is not claimed to be optimal, and the same 300-s setting is used in all comparative experiments without case-wise retuning. The sensitivity results therefore indicate that the principal conclusions are not specific to this exact window length.

For the zero-error controls, one of the three trajectories receives a 0.5-pp correction from 180 s onward. This nominal perturbation, together with the higher zero-error nonzero-correction rates observed for the shorter windows, reinforces the intended conditional use of commissioning: it is activated only when the initial SOC prior is declared uncertain, whereas a reliable initial prior bypasses the commissioning procedure.
\end{revisionblock}

\subsection{Fault accommodation after uncertain-state commissioning}
\label{subsec:delayed_faults}

\begin{revisionblock}
This subsection combines initial-state recovery and subsequent fault accommodation under one controlled representative condition. The 50\%-SOC-anchored $25^\circ$C--DST segment contains 2943 samples at a 1-s sampling interval. The synthetic voltage abnormality is introduced at $t=1472$ s, the first sample after the temporal midpoint. This record-length-based choice gives 1472 healthy pre-fault samples and 1471 post-fault samples, providing both a long post-commissioning healthy interval and a comparably long faulted interval. The onset is fixed independently of estimator behavior and is not supplied to any estimator. The original RBC-DEKF and conventional joint EKF are active from startup, whereas the commissioned RBC-DEKF holds its residual state at zero until the 300 s handover and then remains active.
\end{revisionblock}

\begin{table}[H]
\centering
\color{revisionyellow}
\caption{Representative SOC errors for three abnormalities introduced after uncertain-state commissioning on the same $25^\circ$C--DST trajectory with a $+10$-percentage-point initial-SOC perturbation. RMSE and maximum error are in percentage points; the pre-fault interval begins after the 300 s handover.}
\label{tab:operational_fault_summary}
\small
\setlength{\tabcolsep}{3pt}
\begin{tabular}{>{\raggedright\arraybackslash}p{0.22\textwidth}>{\raggedright\arraybackslash}p{0.27\textwidth}rrr}
\toprule
Fault profile & Method & \makecell{Pre-fault\\RMSE} & \makecell{Post-fault\\RMSE} & \makecell{Post-fault\\maximum} \\
\midrule
$+30$ mV step & Original RBC-DEKF & 9.934 & 9.952 & 10.044 \\
& Conventional joint EKF & 9.952 & 9.942 & 10.011 \\
& Commissioned RBC-DEKF & \textbf{0.656} & \textbf{0.697} & \textbf{0.770} \\
\addlinespace
0--$+30$ mV, 10-s ramp & Original RBC-DEKF & 9.934 & 9.952 & 10.044 \\
& Conventional joint EKF & 9.952 & 9.942 & 10.011 \\
& Commissioned RBC-DEKF & \textbf{0.656} & \textbf{0.697} & \textbf{0.770} \\
\addlinespace
$+2\%$ voltage gain & Original RBC-DEKF & 9.934 & 9.963 & 10.062 \\
& Conventional joint EKF & 9.952 & 9.942 & 10.011 \\
& Commissioned RBC-DEKF & \textbf{0.656} & \textbf{0.715} & \textbf{0.791} \\
\bottomrule
\end{tabular}
\end{table}

\begin{figure}[H]
    \centering
    \includegraphics[width=\textwidth]{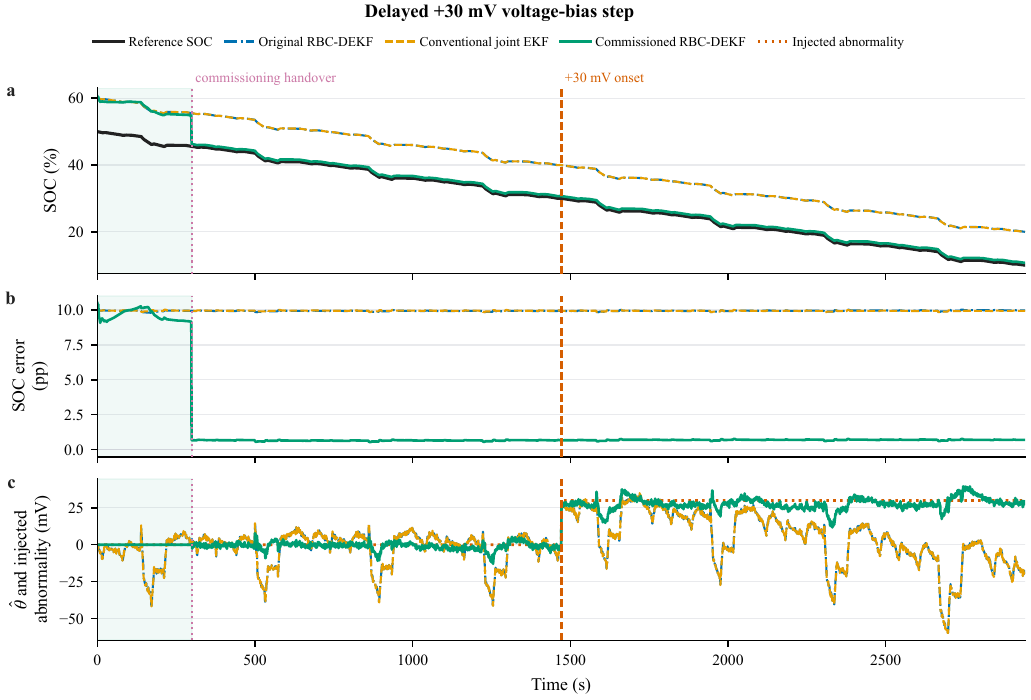}
    \caption{\rev{Representative response to a fault introduced during operation for the $25^\circ$C--DST trajectory. A $+10$-percentage-point initial-SOC perturbation is followed by a delayed $+30$ mV voltage step. (a) SOC estimates; (b) SOC errors; and (c) estimated lumped voltage discrepancy together with the injected fault. The original RBC-DEKF and conventional joint EKF remain active from startup; the commissioned RBC-DEKF suspends its residual state until the 300 s handover and does not receive the later fault-onset time.}}
    \label{fig:initial_then_fault}
\end{figure}

\begin{figure}[H]
    \centering
    \includegraphics[width=\textwidth]{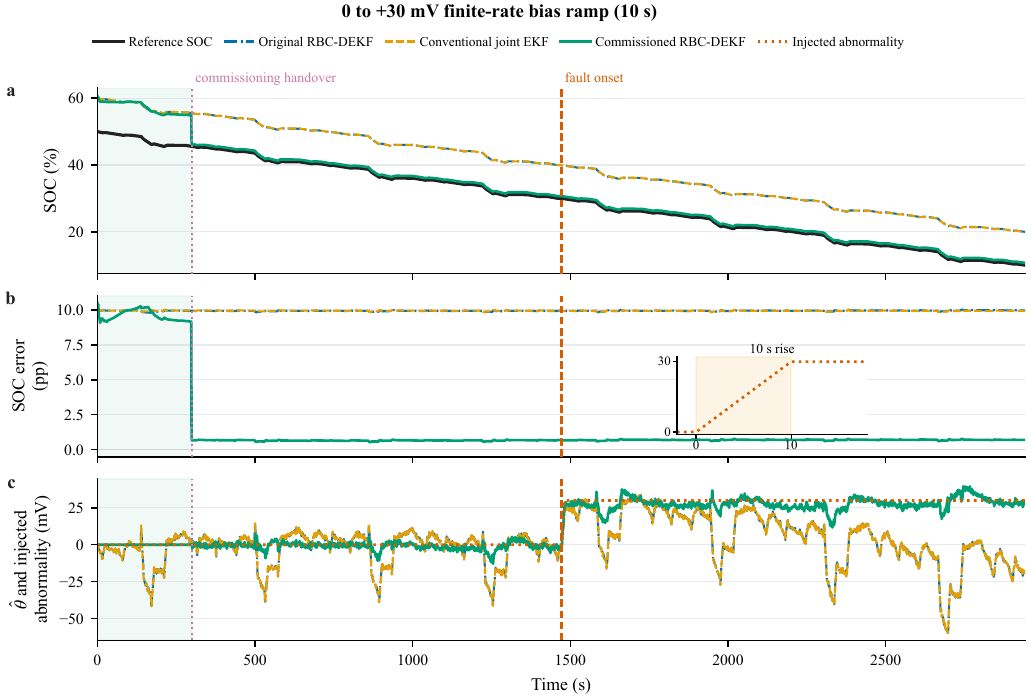}
    \caption{\rev{Representative response to a voltage-bias ramp on the $25^\circ$C--DST trajectory with a $+10$-percentage-point initial-SOC perturbation. The injected offset rises from 0 to $+30$ mV over 10 s. Panels (a)--(c) show SOC, SOC error, and the estimated lumped voltage-discrepancy state together with the injected abnormality; the inset resolves the prescribed 10-s transition. The commissioned RBC-DEKF activates its residual loop after the 300 s handover, and no estimator receives the fault-onset time.}}
    \label{fig:dynamic_ramp}
\end{figure}

\begin{figure}[H]
    \centering
    \includegraphics[width=\textwidth]{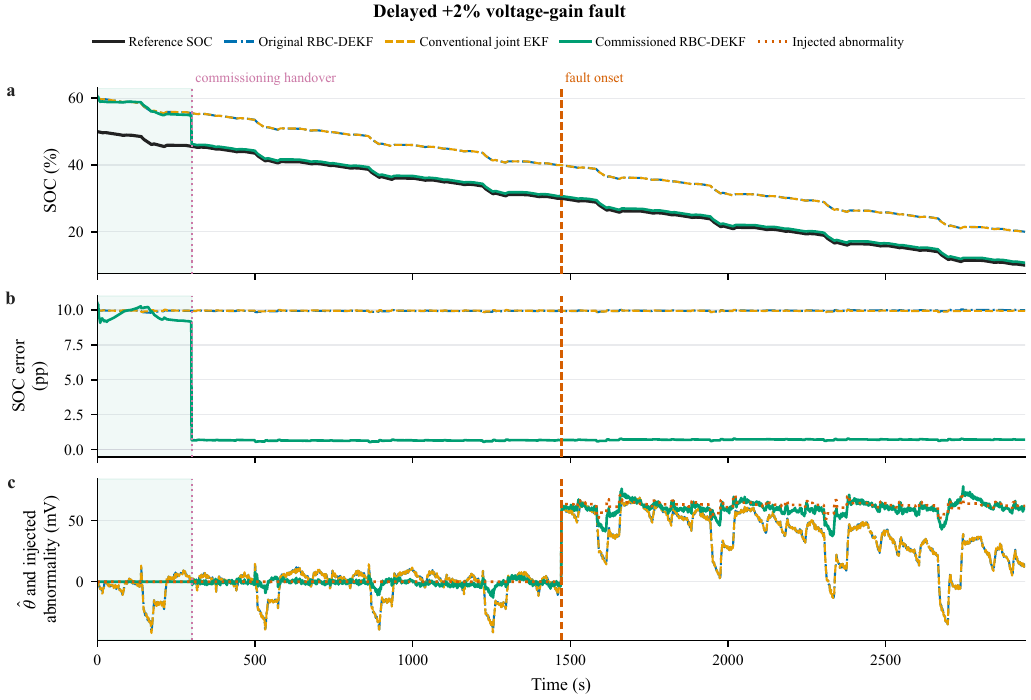}
    \caption{\rev{Representative response to a delayed $+2\%$ voltage-gain fault on the same $25^\circ$C--DST trajectory and initial condition. Panels (a)--(c) show SOC, SOC error, and the estimated lumped voltage-discrepancy state together with the injected equivalent voltage term $0.02V_k^{\mathrm{true}}$. The commissioned RBC-DEKF activates its residual loop after the 300 s handover, and no estimator receives the fault-onset time.}}
    \label{fig:dynamic_gain}
\end{figure}

\begin{revisionblock}
Table~\ref{tab:operational_fault_summary} provides a common numerical basis for Figures~\ref{fig:initial_then_fault}--\ref{fig:dynamic_gain}. Across all three profiles, the original and joint estimators preserve nearly all of the startup SOC offset, whereas the commissioned RBC-DEKF first corrects that offset and then maintains a post-fault SOC RMSE of 0.697--0.715 percentage points, descriptively 0.041--0.059 percentage points above the pre-fault value. These differences compare distinct time intervals and are not isolated causal estimates of fault-induced degradation. The trajectories nevertheless show that recovered SOC remains stable after the abnormalities are introduced.
\end{revisionblock}

\section{Discussion}
\label{sec:discussion}

\subsection{Why a dual realization: functional decoupling versus joint augmentation}

\begin{revisionblock}
The conventional joint-EKF baseline propagates the full cross covariance and is more accurate on average than the original RBC-DEKF under accurate initialization and the inherited covariance values (0.092811 versus 0.251308 percentage points over all 384 static condition--fault profiles).

Under accurate initialization, when electrochemical-state errors are small, the remaining voltage innovation can be dominated by the lumped voltage discrepancy rather than a genuine SOC mismatch. In this regime, the Joint-EKF uses the propagated residual-state uncertainty and state--residual cross covariance within one common augmented update [Eqs.~\eqref{eq:joint_innovation_cov_expanded}--\eqref{eq:joint_gain_blocks}], which can moderate unnecessary electrochemical-state correction. By contrast, the sequential RBC-DEKF first updates the electrochemical state using a gain that does not directly contain $P_{\theta,k}^{-}$ [Eq.~\eqref{eq:dual_covariance_decoupling}], and only afterward updates the voltage-discrepancy state using the same voltage sample. Consequently, part of a discrepancy-dominated innovation can enter the electrochemical-state correction before the residual loop adapts. This provides a first-order mechanism consistent with the lower Joint-EKF SOC RMSE observed under accurate initialization. Because the two realizations also use different measurement-weighting structures and inherited covariance tuning, the observed performance difference is not attributed to cross-covariance propagation alone.

The dual realization instead maintains separate state and residual design covariances, so $P_{\theta,k}^-$ does not directly change $K_{x,k}^D$ and the scalar residual recursion can be suspended while the four-state Joseph covariance update is retained. This functional decoupling enables the commissioning procedure while retaining the four-state EKF recursion structure. In the representative $+10$-percentage-point case of Figure~\ref{fig:commissioning_mechanism}, the two always-on residual-state estimators retain errors near 10 percentage points, whereas the complete commissioned realization attains a post-handover SOC RMSE of 0.671 percentage points. The result supports the combined startup procedure---including bounded state correction and frozen healthy-residual calibration---rather than a universal numerical ranking of dual and joint filters.
\end{revisionblock}

\subsection{Initial-state uncertainty and intrinsic observability}

\begin{revisionblock}
On the flat plateau, the local startup innovation has the schematic form
\begin{equation}
\nu_k\simeq h_s\,\delta\mathrm{SOC}_k+\delta\theta_k+v_k,
\label{eq:soc_residual_ambiguity}
\end{equation}
with small $|h_s|$. Adding $\theta$ does not increase $|h_s|$, add a sensor, or create intrinsic SOC observability. It instead adds another explanation for the same innovation and can therefore preserve an initial SOC error when the residual loop adapts rapidly. Commissioning temporarily imposes the verified-healthy constraint $\theta=0$ and restricts the startup correction to the model-generated, SOC-aligned trajectories defined in Section~\ref{subsec:commissioning}. The window objective combines these trajectories with the disclosed frozen healthy-residual calibration; it does not add an online measurement channel or guarantee unique physical-state recovery. If a voltage fault is already present at sample zero, or if the healthy model discrepancy is inadequately calibrated, the two contributions remain non-identifiable from this single voltage channel. The proposed procedure must therefore be viewed as conditional state recovery under a healthy-window assumption.

The commissioning correction acts on the electrochemical state coordinates rather than directly resetting scalar SOC. Equation~\eqref{eq:candidate_electrochemical_initial_state} follows the two-electrode stoichiometric mapping of Eq.~\eqref{eq:initial_state_mapping}; in the 50\%-SOC-anchored tests, the initial perturbation and candidate correction use the same SOC-aligned direction while retaining the checkpoint's pre-existing dynamic profile information. Each candidate is propagated over the measured current--temperature window and ranked by the full calibrated terminal-voltage prediction, which combines both electrode OCPs with diffusion, kinetic, and ohmic terms. The result is a paired-electrode, model-constrained initialization rather than inversion of a scalar OCV--SOC curve.

Although the quasi-equilibrium full-cell OCV slope is small on the LFP plateau, the commissioning objective uses a finite current-driven trajectory rather than a single voltage sample. SOC-aligned candidates generate different paired-electrode concentration trajectories, whose concentration-dependent OCP and kinetic contributions can produce different terminal-voltage predictions over the buffered window. Accumulating these voltage-consistency differences helps rank candidate initial states under the verified-healthy assumption. The available trajectory and model calibration determine the usefulness of this comparison; the procedure does not create new instantaneous SOC observability.
\end{revisionblock}

\subsection{Equivalent-residual interpretation of multiplicative gain faults}
\label{subsec:gain_discussion}

The gain-fault results can be interpreted most clearly in the voltage-residual domain. For a constant gain coefficient $\alpha$, the induced residual is
\begin{equation}
b_k^{\mathrm{gain}}=\alpha V_k^{\mathrm{true}}.
\label{eq:gain_equivalent_residual}
\end{equation}
By decomposing the true terminal voltage as $V_k^{\mathrm{true}}=\overline{V}^{\mathrm{true}}+\widetilde{V}_k^{\mathrm{true}}$, the equivalent residual becomes
\begin{equation}
b_k^{\mathrm{gain}}
=
\alpha\overline{V}^{\mathrm{true}}
+
\alpha\widetilde{V}_k^{\mathrm{true}}.
\label{eq:gain_residual_decomposition}
\end{equation}
The first term represents the mean voltage-domain offset, whereas the second describes its variation with terminal voltage. The voltage-discrepancy state estimates the combined systematic voltage discrepancy, including the gain-induced residual and model mismatch. Therefore, although the RBC-DEKF does not identify the gain coefficient $\alpha$ explicitly, its scalar residual state can represent the instantaneous voltage-domain effect of the gain fault.

\rev{The matched-magnitude results in Section~\ref{subsec:overall_results} show similar accommodation of additive and gain faults when their equivalent voltage residuals are comparable. Remaining errors are concentrated where the gain-induced residual is larger and more time-varying than the tested additive offsets. The observed sign asymmetry indicates interaction with nonlinear voltage--state sensitivity and fixed covariance tuning. Explicit bias--gain estimation may be useful for fault-mode identification, while the present scalar state supports the reported voltage-domain accommodation.}

\subsection{Implications for flat-plateau battery chemistries}
\label{subsec:flat_plateau_implications}

The results demonstrate that fault-tolerant state estimation cannot be considered independently of battery chemistry. In LFP batteries, the two-phase plateau weakens the mapping from terminal voltage to lithiation state, allowing systematic voltage abnormalities to dominate the voltage innovation used by a voltage-corrected observer. The RBC-DEKF provides a separate adaptive channel for the dominant systematic voltage discrepancy alongside the electrochemical state estimator.

The underlying design principle may also be relevant to other flat-voltage or phase-transition-dominated electrode systems. However, the quantitative compensation performance remains chemistry dependent because it is governed by the local voltage--composition slope, dynamic polarization, hysteresis, and model fidelity. Extension to other chemistries therefore requires chemistry-specific parameterization and independent validation rather than direct transfer of the numerical results reported here.

\Needspace{5\baselineskip}
\subsection{Model mismatch, hysteresis, and residual interpretation}
\label{subsec:residual_interpretation}

\begin{revisionblock}
The estimated residual must be interpreted as
\begin{equation}
\theta_k=\theta_{\mathrm{sensor},k}+\delta_{\mathrm{model},k}
+\delta_{\mathrm{hys},k}+\delta_{\mathrm{other},k},
\label{eq:lumped_residual_sources}
\end{equation}
not as a source-isolated sensor bias. Averaging low-rate charge and discharge OCV branches reduces first-order branch separation in the material-sensitivity plot, but does not remove dynamic hysteresis. The CPG-SPMT represents diffusion, charge-transfer polarization, ohmic loss, and temperature-dependent grouped parameters; it does not contain an explicit hysteresis state. Hysteresis, relaxation not captured by the reduced dynamics, parameter error, and voltage-sensor abnormality can consequently enter the same scalar residual.

This distinction also changes the interpretation of voltage RMSE. A small posterior measurement-fit error shows that the augmented output can reconstruct the faulty sample after using that sample for correction. It is not evidence that the clean terminal voltage or physical sensor bias has been recovered. Source isolation would require additional information, such as redundant sensors, an explicit validated hysteresis model, fault-specific dynamics, or an independently calibrated healthy-residual envelope.
\end{revisionblock}

\subsection{Practical implications and future extensions}
\label{subsec:future_work}

\rev{From a battery-management perspective, the results show that a one-state voltage-discrepancy loop can improve the robustness of electrochemical-model-based SOC estimation under the tested additive-bias and multiplicative-gain voltage abnormalities without relying on sensor redundancy or SOC-region partitioning. This is particularly relevant to LFP batteries, for which the flat OCV response amplifies small voltage errors into large SOC ambiguity. The online voltage-discrepancy recursion adds only one scalar state; the uncertain-start branch additionally uses the disclosed offline calibration and finite-window commissioning procedure. Explicit fault-parameter identification is not required for the reported voltage-domain accommodation; embedded resource use remains to be measured.}

\rev{The explicit estimate $\hat{\theta}_k$ may provide a candidate online abnormality signal. Because the estimated quantity is lumped, however, it does not by itself establish a sensor fault, isolate sensor effects from model mismatch, or classify additive and multiplicative modes. Reliable diagnostic decisions require a separately validated healthy residual baseline, adaptive thresholds, or an additional classification layer. A further limitation is that the present validation uses synthetically imposed measurement abnormalities on an open-source dataset; dedicated hardware experiments are required to quantify actual sensor drift, calibration errors, communication latency, and embedded implementation constraints.}

\begin{revisionblock}
The present evidence has five practical boundaries. First, the voltage abnormalities are synthetic, the delayed profiles are representative accommodation tests rather than a detection benchmark, and autonomous fault classification is not validated. Second, the model and estimator assume a fixed health state. Aging, capacity fade, resistance growth, and state-of-health adaptation are outside the reported results. Third, current-sensor bias and joint current/voltage faults are not evaluated. Fourth, the commissioning study uses three representative 50\%-SOC-anchored trajectories, one noise realization, and a supervised cross-cycle healthy-residual map; its repeated initial offsets are paired sensitivity cases rather than independent cells. Fifth, because all 24 temperature--cycle trajectories participate in parameter identification and/or candidate selection, the full static-fault matrix measures in-matrix operating-condition coverage rather than out-of-sample generalization.

Future work should therefore combine explicit hysteresis and aging-aware electrochemical dynamics with capacity/SOH estimation; assess parameter mismatch and current-sensor bias; validate the uncertain-start commissioning decision on independent healthy cells; develop and validate a separate sensor-fault decision layer; and test physical sensor faults in hardware-in-the-loop and embedded BMS implementations. Joint bias--gain states may also be useful when direct fault-mode identification is required rather than voltage-domain accommodation alone.
\end{revisionblock}

\section{Conclusion}
\begin{revisionblock}
This work developed an electrochemically constrained RBC-DEKF for accurate LFP SOC estimation under abnormal voltage measurements and uncertain initialization. The CPG-SPMT supplies paired-electrode concentration dynamics, while a separately configurable scalar residual loop accommodates the lumped systematic voltage discrepancy. Across 24 temperature--drive-cycle trajectories, the RBC-DEKF reduced mean SOC RMSE from 7.646 to 0.170 percentage points over 216 additive-bias profiles and from 22.646 to 0.356 percentage points over 168 multiplicative-gain profiles.

For uncertain initialization, commissioning suspends the residual estimator, corrects paired-electrode initial states using buffered healthy data, and replays the calibrated state EKF before handover. Across 18 paired nonzero-initial-error profiles on three 50\%-SOC-anchored segments, full-record mean SOC RMSE was 1.747 pp, including the startup transient, compared with 11.192 and 11.400 pp for the always-on RBC-DEKF and conventional joint EKF. Correction accuracy was stable over the tested 180--600-s windows, supporting the retained 300-s setting. Representative delayed step, 10-s ramp, and gain tests yielded post-fault RMSE values of 0.697--0.715 pp.

The central novelty is the electrochemically constrained temporal coordination of uncertain-state recovery and subsequent voltage-discrepancy accommodation: the residual channel is suspended while an SOC-aligned, physically consistent paired-electrode state is selected from a verified-healthy trajectory, after which residual adaptation is reactivated for fault-tolerant operation. The reported commissioned accuracy belongs to the complete realization, including the frozen healthy-model calibration. Physical sensor-fault validation, aging-aware calibration, and fault-source classification remain directions for further work.
\end{revisionblock}

\appendix
\setcounter{table}{0}
\renewcommand{\thetable}{A.\arabic{table}}
\renewcommand{\theHtable}{A.\arabic{table}}
\setcounter{figure}{0}
\renewcommand{\thefigure}{A.\arabic{figure}}
\renewcommand{\theHfigure}{A.\arabic{figure}}
\Needspace{7\baselineskip}
\section{LFP-specific parameter identification}
\label{app:parameter_identification}

\begin{revisionblock}
\subsection{Data matrix and parameterization}

The CPG-SPMT parameterization was identified from the complete A123 operating-condition matrix: DST, FUDS, and US06 at each of $-10$, 0, 10, 20, 25, 30, 40, and $50^\circ$C, for a total of 24 measured trajectories. The parameter vector is
\begin{equation}
\mathbf p=\big[\alpha_{n,1},\alpha_{p,1},b_n,b_p,d_{n,1},d_{p,1},
s_{n,0},s_{p,0},E_1,E_2,E_3,E_4,E_5,R_{\mathrm{ini},1}\big]^\top.
\label{eq:identification_vector}
\end{equation}
The five activation energies couple the reference transport, kinetic, and ohmic parameters across temperature through the Arrhenius relations in Section~2.2. The positive- and negative-electrode OCP expressions follow the CPG-SPMT formulation in Ref.~\cite{GUO2026110075}.

\subsection{Bound-constrained multi-start optimization}

The optimization workflow follows the batch-optimized local-to-global concept in Ref.~\cite{guo2026bolt}. Initial candidates are sampled independently within the prescribed box $[\mathbf l,\mathbf u]$. Each candidate is refined using a bound-constrained trust-region reflective least-squares solver. To avoid giving longer trajectories greater weight solely because they contain more samples, the refinement objective uses an equal-condition mean,
\begin{align}
r_{m,k}(\mathbf p)&=V_{m,k}^{\mathrm{meas}}-
V_{m,k}^{\mathrm{CPG-SPMT}}(\mathbf p),\\
J_{\mathrm{LS}}(\mathbf p)&=
\frac{1}{24}\sum_{m=1}^{24}\frac{1}{N_m}
\sum_{k=1}^{N_m}r_{m,k}^2(\mathbf p),
\qquad \mathbf l\leq\mathbf p\leq\mathbf u.
\label{eq:parameter_ls_objective}
\end{align}
The refined candidates are ranked by the unweighted mean of condition-wise voltage MAEs,
\begin{align}
\mathrm{MAE}_m^{(i)}&=\frac{1}{N_m}\sum_{k=1}^{N_m}
\left|r_{m,k}(\hat{\mathbf p}^{(i)})\right|,\\
\mathbf p^*&=\arg\min_i\frac{1}{24}\sum_{m=1}^{24}
\mathrm{MAE}_m^{(i)}.
\label{eq:parameter_candidate_selection}
\end{align}
This strategy makes the numerical optimizer chemistry-independent, but the model output, OCP functions, stoichiometric windows, bounds, and selected parameter vector remain LFP-specific. Multi-start refinement improves practical repeatability but does not prove global convergence or parameter uniqueness.
\end{revisionblock}

\begin{table}[H]
\centering
\color{revisionyellow}
\caption{Bounds supplied for the 14-parameter A123 LFP/graphite identification and selected values used in this study.}
\label{tab:parameter_bounds}
\scriptsize
\setlength{\tabcolsep}{4pt}
\begin{tabular}{lrrrl}
\toprule
Parameter & Lower bound & Upper bound & Selected value & Unit \\
\midrule
$\alpha_{n,1}$ & 1 & 4000 & 2696.83 & s \\
$\alpha_{p,1}$ & 1 & 2000 & 1352.43 & s \\
$b_n$ & 5000 & 11000 & 10999.93 & C \\
$b_p$ & 3600 & 9000 & 3982.26 & C \\
$d_{n,1}$ & $1\times10^{-4}$ & $1\times10^{-2}$ & $5.475\times10^{-3}$ & s mol$^{1/2}$ m$^{-5/2}$ \\
$d_{p,1}$ & $1\times10^{-5}$ & $1\times10^{-3}$ & $3.43\times10^{-4}$ & s mol$^{1/2}$ m$^{-5/2}$ \\
$s_{n,0}$ & 0.3 & 0.7 & 0.3829 & -- \\
$s_{p,0}$ & $1\times10^{-10}$ & 0.01 & $2.82\times10^{-4}$ & -- \\
$E_1$ & $1\times10^{4}$ & $1\times10^{5}$ & 44042.22 & J mol$^{-1}$ \\
$E_2$ & $1\times10^{4}$ & $1\times10^{5}$ & 52664.43 & J mol$^{-1}$ \\
$E_3$ & $1\times10^{3}$ & $1\times10^{6}$ & 133074.86 & J mol$^{-1}$ \\
$E_4$ & $1\times10^{4}$ & $1\times10^{5}$ & 14397.71 & J mol$^{-1}$ \\
$E_5$ & $1\times10^{3}$ & $1\times10^{5}$ & 4097.71 & J mol$^{-1}$ \\
$R_{\mathrm{ini},1}$ & 0.15 & 0.20 & 0.1733 & $\Omega$ \\
\bottomrule
\end{tabular}
\end{table}

\begin{revisionblock}
All selected values lie inside the specified box. The value $b_n=10999.93$ C is close to its upper bound of 11000 C, so it should not be interpreted as a tightly isolated physical parameter without additional excitation or a profile-likelihood analysis. Because all 24 measured trajectories enter refinement and/or candidate selection, the measured-data fit represents cross-condition consistency over the available matrix; no trajectory is claimed as an independent held-out model-validation set. The fault-injection results that reuse these trajectories test estimator behavior under controlled measurement perturbations, not out-of-sample parameter generalization.
\end{revisionblock}

\subsection{SOC-dependent raw voltage-residual audit}
\label{app:soc_residual_audit}

\begin{figure}[H]
    \centering
    \includegraphics[width=\textwidth]{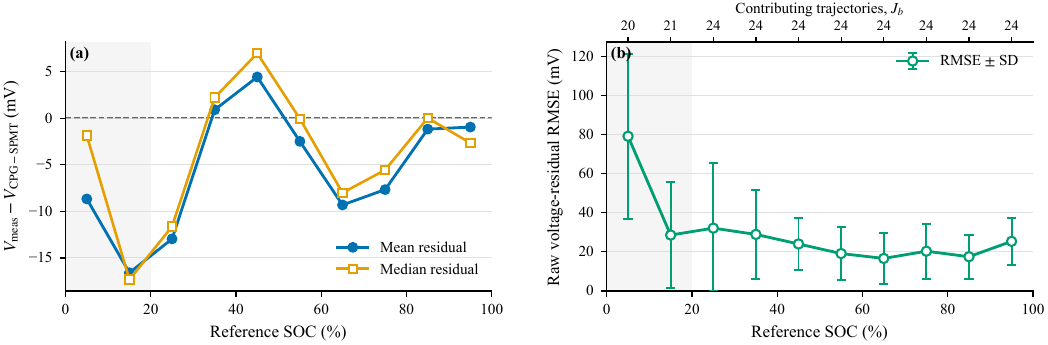}
    \caption{\rev{Raw CPG-SPMT voltage residual versus coulomb-counted reference SOC over the eight temperatures and three drive cycles. (a) Equal-weight averages of the within-trajectory mean and median signed residual. (b) Equal-weight mean RMSE with descriptive inter-trajectory standard deviation; lower whiskers are truncated at zero. The labels above panel (b) give the number $J_b$ of contributing temperature--cycle trajectories; the $0$--$10\%$ and $10$--$20\%$ bins are lightly shaded because fewer than 24 records reach those intervals.}}
    \label{fig:soc_residual_by_soc}
\end{figure}

\begin{revisionblock}
To resolve the SOC-dependence part of the model-error query, the raw residual was recomputed at every sample of all 24 complete temperature--cycle records as
\begin{equation}
r_{\mathrm{model},k}=V_k^{\mathrm{meas}}-V_k^{\mathrm{CPG-SPMT}}.
\label{eq:raw_model_residual}
\end{equation}
The reference SOC is obtained by coulomb counting from the 100\% record initialization using the measured current and the stated 1.0355-Ah capacity. Samples are assigned to ten left-closed SOC intervals, with the last interval including 100\%. Statistics are first computed within each trajectory and SOC bin and are then averaged equally across contributing trajectories, so long records do not dominate the result. The error bars in Figure~\ref{fig:soc_residual_by_soc} are the descriptive sample standard deviation of the trajectory-wise RMSE values, not confidence intervals; lower whiskers are truncated at zero because RMSE is nonnegative. Coverage is incomplete below 20\% SOC because several low-temperature records terminate earlier; the contributing-trajectory counts are therefore shown explicitly.
\end{revisionblock}

\clearpage
\setcounter{table}{0}
\renewcommand{\thetable}{B.\arabic{table}}
\renewcommand{\theHtable}{B.\arabic{table}}
\section{Healthy-residual calibration and component ablation}
\label{app:commissioning_ablation}

\begin{revisionblock}
\subsection{Calibration implementation}
\label{app:calibration_implementation}

The auxiliary calibration represents repeatable voltage discrepancy of the healthy CPG-SPMT and is trained before estimator operation, independently of the online residual state. For a target drive cycle $c$ at nominal temperature $T_{\mathrm{cond}}$, the regressor is fitted to healthy samples from the other two drive cycles at the same temperature. The training target, expressed in mV, is
\begin{equation}
r_{h,j}^{\mathrm{mV}}=10^3\!\left[V_{\mathrm{meas},j}
-h\!\left(\mathbf x_j^{\mathrm{OL}},I_j,T_j\right)\right],
\label{eq:healthy_residual_target}
\end{equation}
where $\mathbf x_j^{\mathrm{OL}}$ is generated by open-loop CPG-SPMT propagation. With $s_j^{\mathrm{pct}}$ denoting SOC in percentage units, the feature vector is
\begin{equation}
\boldsymbol\psi_j=\left[
s_j^{\mathrm{pct}}/100,\ I_j,\ |I_j|,\ I_j^2,\
\bar I_j^{(0.10)},\ \bar I_j^{(0.02)},\ \Delta I_j
\right]^{\!\top},
\label{eq:healthy_residual_features}
\end{equation}
where $\Delta I_j=I_j-I_{j-1}$ and each filtered-current feature follows $\bar I_j^{(a)}=aI_j+(1-a)\bar I_{j-1}^{(a)}$, initialized from the first current sample. A histogram-gradient-boosting regressor is fitted with 200 iterations, at most 31 leaf nodes, $L_2$ regularization of 1, and random seed 1.

For the held-out target current trajectory, the fitted model is evaluated at every sample on an SOC grid from 0 to 100\% with 1-percentage-point spacing. The lookup coordinate $s^{\mathrm{pct}}$ is expressed in percentage points. This creates a sample-by-SOC table,
\begin{equation}
\hat\delta_{h,k}^{(T_{\mathrm{cond}},-c)}(s^{\mathrm{pct}})
=10^{-3}f_{T_{\mathrm{cond}}}^{(-c)}\!\left(\boldsymbol\psi_k(s^{\mathrm{pct}})\right),
\label{eq:healthy_residual_lookup}
\end{equation}
which is linearly interpolated at the candidate or estimated SOC. The current and current-history features use the measured target-cycle current through sample $k$; only the SOC coordinate is swept across the lookup grid. Coulomb-counted reference SOC is used only in offline fitting and the held-out audit. During estimator operation, the frozen table is queried using model-estimated SOC. Thus, $\hat\delta_h$ is a fixed healthy-model calibration, distinct from the adaptive online voltage-discrepancy state $\hat\theta$.

The derivative used in Eq.~\eqref{eq:calibrated_jacobian} is the slope between neighboring 1-pp SOC grid entries multiplied by the paired-electrode SOC Jacobian. At an interior grid knot the implementation selects the right interval. Outside the lookup range, the map value is held at the nearest endpoint; the implementation retains the nearest interior slope as an approximate EKF linearization, not the exact derivative of the clipped lookup. The exact derivative of the constant extension would be zero. The SOC Jacobian uses the affine two-electrode mapping. These conventions specify the implemented gain calculation as well as the predicted voltage. Precomputing a table for the recorded current trace is an implementation convenience: the features at sample $k$ use only current measurements through $k$, so each row can also be constructed causally.

\subsection{Calibration contribution}
\label{app:calibration_ablation}

The two configurations below differ only in healthy-residual calibration. Both use the same residual suspension, bounded ordinary-$L_2$ correction, causal replay, and state-first handover, with identical state-prior and residual-loop settings. RMSE covers the complete recorded output, including startup, over the same 18 nonzero-initial-error profiles as Table~\ref{tab:initial_soc_recovery}.
\end{revisionblock}

\begin{table}[H]
\centering
\color{revisionyellow}
\caption{Calibration ablation using full-record arithmetic-mean SOC RMSE over 18 paired nonzero-initial-error profiles. Startup provisional outputs are included.}
\label{tab:commissioning_calibration_ablation}
\small
\begin{tabular}{>{\raggedright\arraybackslash}p{0.72\textwidth}r}
\toprule
Commissioning configuration & SOC RMSE (pp) \\
\midrule
Three-stage commissioning without healthy-residual calibration & 8.665 \\
Three-stage commissioning with cross-cycle healthy-residual calibration & 1.747 \\
\bottomrule
\end{tabular}
\end{table}

\begin{revisionblock}
The calibration reduces mean SOC RMSE from 8.665 to 1.747 pp. This difference quantifies its contribution to the complete commissioned realization; suspension and trajectory correction are retained in both rows and are not isolated by this ablation.

This large difference is consistent with the calibration's role in trajectory matching. On the flat LFP plateau, a repeatable healthy-model voltage bias can otherwise be explained by shifting candidate SOC, distorting the ranking of initial electrochemical states. Accounting for this repeatable component in the voltage prediction reduces its influence on candidate ranking. After handover, the online residual state remains available to accommodate deviations beyond the frozen healthy baseline. Because calibration also enters replay and subsequent updates, the full-record ablation measures its combined contribution, not the search-stage effect alone.
\end{revisionblock}

\section*{Funding}
This work was supported by the Research Foundation -- Flanders (FWO) under grant number 1252326N.

\section*{Declaration of competing interest}
The authors declare that they have no known competing financial interests or personal relationships that could have appeared to influence the work reported in this paper.

\section*{Data and code availability}
The experimental data used in this study are publicly available from the CALCE Battery Research Group (\url{https://calce.umd.edu/battery-data#A123}). The CPG-SPMT electrochemical-model implementation is available at \url{https://github.com/FrankSuperG/CPG-SPMT}.

\section*{CRediT authorship contribution statement}
\textbf{Feng Guo:} Conceptualization, Methodology, Software, Validation, Formal analysis, Investigation, Data curation, Visualization, Writing -- original draft, Writing -- review \& editing.
\textbf{Luis D. Couto:} Methodology, Writing -- review \& editing.
\textbf{Hamid Hamed:} Methodology, Writing -- review \& editing.
\textbf{Khiem Trad:} Methodology, Writing -- review \& editing.
\textbf{Dong Zhang:} Writing -- review \& editing.
\textbf{Ru Hong:} Writing -- review \& editing.
\textbf{Guangdi Hu:} Writing -- review \& editing.
\textbf{Mohammadhosein Safari:} Supervision, Resources, Writing -- review \& editing.

\bibliographystyle{elsarticle-num}
\bibliography{references}

\end{document}